# Privy: Envisioning and Mitigating Privacy Risks for Consumer-facing AI Product Concepts

Hao-Ping (Hank) Lee
haopingl@cs.cmu.edu
Carnegie Mellon University
Pittsburgh, PA, United States

Yu-Ju Yang*
yuju2@illinois.edu
University of Illinois
Urbana-Champaign
Champaign, IL, United States

Matthew Bilik*
bilik@uw.edu
University of Washington
Seattle, WA, United States

Isadora Krsek*
ikrsek@andrew.cmu.edu
Carnegie Mellon University
Pittsburgh, PA, United States

Thomas Serban von Davier
tvondavi@andrew.cmu.edu
Carnegie Mellon University
Pittsburgh, United States

Kyzyl Monteiro
kmonteir@andrew.cmu.edu
Carnegie Mellon University
Pittsburgh, PA, United States

Jason Lin
jasonlin562@gmail.com
Carnegie Mellon University
Pittsburgh, PA, United States

Shivani Agarwal
shivania@andrew.cmu.edu
Carnegie Mellon University
Pittsburgh, PA, United States

Jodi Forlizzi
forlizzi@cs.cmu.edu
Carnegie Mellon University
Pittsburgh, PA, United States

Sauvik Das
sauvik@cmu.edu
Carnegie Mellon University
Pittsburgh, PA, United States

## Abstract

AI creates and exacerbates privacy risks, yet practitioners lack effective resources to identify and mitigate these risks. We present Privy, a tool that guides practitioners without privacy expertise through structured privacy impact assessments to: (i) identify relevant risks in novel AI product concepts, and (ii) propose appropriate mitigations. Privy was shaped by a formative study with 11 practitioners, which informed two versions — one LLM-powered, the other template-based. We evaluated these two versions of Privy through a between-subjects, controlled study with 24 separate practitioners, whose assessments were reviewed by 13 independent privacy experts. Results show that Privy helps practitioners produce privacy assessments that experts deemed high quality: practitioners identified relevant risks and proposed appropriate mitigation strategies. These effects were augmented in the LLM-powered version. Practitioners themselves rated Privy as being useful and usable, and their feedback illustrates how it helps overcome long-standing awareness, motivation, and ability barriers in privacy work.

*These authors contributed equally to this research.

## 1 Introduction

Practitioners building AI products often lack awareness of how privacy risks manifest, are not incentivized to address them, and have few resources to guide effective mitigation [38]. Although prior work has proposed taxonomies to characterize emerging privacy threats in AI systems [40, 61, 72], and broad guidelines for building responsible AI products that address some privacy concerns [7, 8, 47], there remains a critical tooling gap. Practitioners often lack both privacy expertise and the support needed to apply these frameworks to early stage product development. This reflects the broader "principle-practice gap" in ethical and responsible AI — guidelines are plentiful but rarely translate into actionable design processes [58, 73]. In this paper, we ask: **How can we design tools**

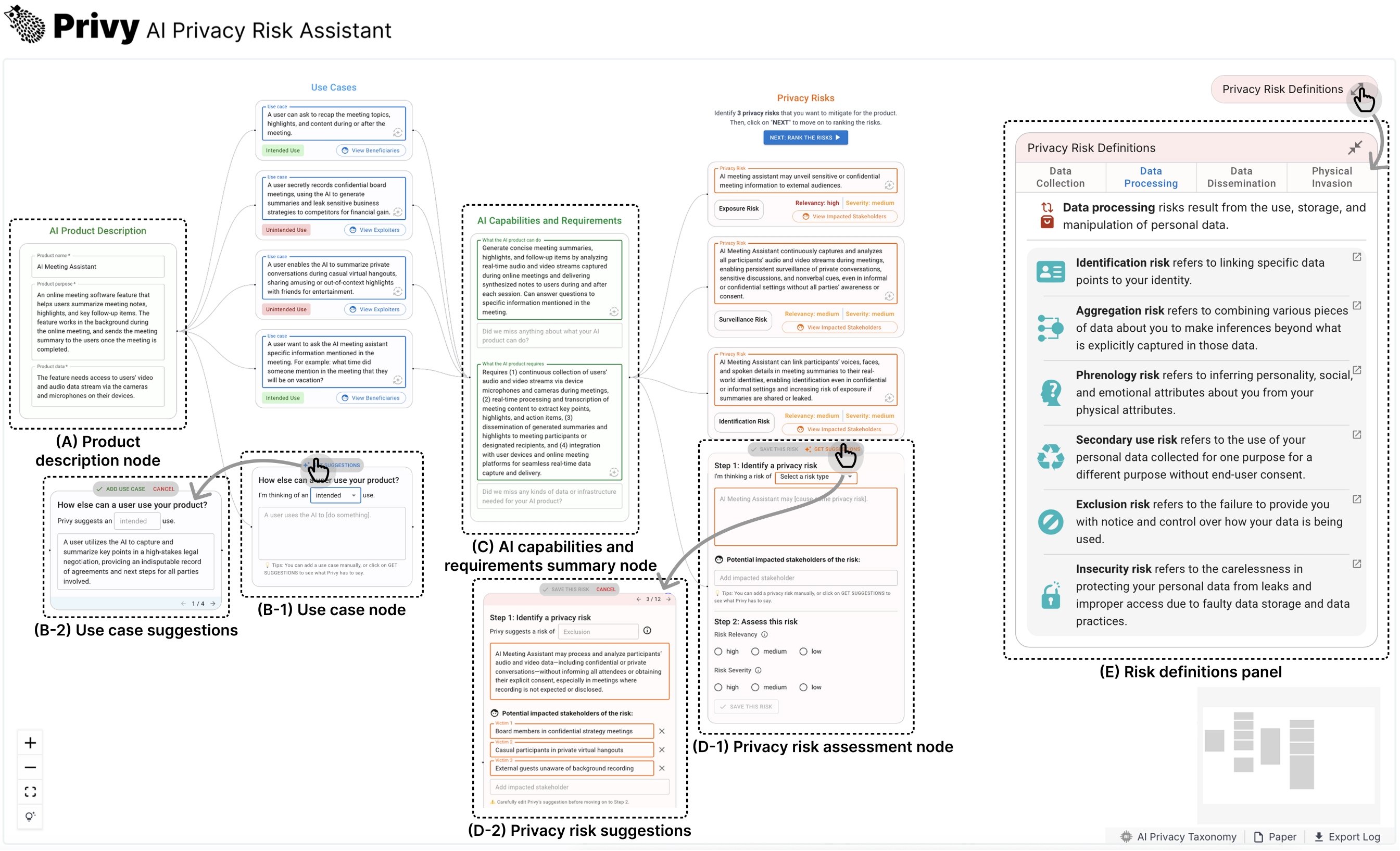

**Figure 1: AI Capability & Requirement Scaffolder: users (A) draft a description of their AI product concept, (B-1) brainstorm *intended* and *unintended* use cases, (B-2) receive suggestions of potential intended and unintended use cases by clicking on "GET SUGGESTIONS", and (C) summarize AI capabilities and requirements of a product concept based on its envisioned use cases. Privacy Risk Explorer: users (D-1) envision how a risk may arise in their product and who may be impacted, rate its *relevancy* and *severity*, and (D-2) receive suggestions of potential privacy risks entailed by their product by clicking on "GET SUGGESTIONS". The (E) *risk definition node* presents definitions of each risk type from the AI privacy taxonomy [40], with external links to real-world incidents.**

**that make it easier for practitioners to identify and mitigate privacy risks in their AI products and services?**

Canonically, privacy experts conduct "privacy impact assessments" (PIAs) to identify privacy risks with novel product concepts [16, 74]. PIAs take the form of a structured report that prompts experts to think about how the design of a product may cause privacy risks, and how those risks might be mitigated [2, 65]. However, existing PIA frameworks (e.g., ISO's Information Technology Security Techniques [2]) have not been substantially updated to account for the evolving and emerging privacy risks introduced by AI technologies [40]. Moreover, preparing these reports requires strong privacy expertise and is labor-intensive, precluding participation by contributing practitioners who are *not* themselves privacy experts.

More recently, researchers have explored the use of Large Language Model (LLM) technologies to help practitioners surface ethical concerns and harms [14, 34, 68]. However, no work has produced LLM-powered systems that might support practitioners in both identifying and mitigating *privacy* risks. Building off this prior work, we developed Privy, a privacy risk-envisioning tool that helps AI practitioners — even those with little privacy expertise — identify and mitigate AI privacy risks in novel product concepts.

To inform the design of Privy, we first implemented an early prototype by adapting Farsight, a general-purpose AI risk envisioning workflow [68] from the responsible AI literature, to the privacy domain using the AI privacy risk taxonomy [40]. With this early prototype, we conducted a formative study with 11 AI and privacy practitioners to distill key design goals. In short, Privy guides AI practitioners through a structured workflow to: (i) elicit the capabilities and requirements of their AI product concepts (Fig. 1A-C), (ii) identify product-specific privacy risks (Fig. 1D), and (iii) brainstorm pertinent mitigation strategies (Fig. 3). We implemented two versions of Privy: 1) Privy$_{Template}$: a non-AI, static template-based version, and 2) Privy$_{LLM}$: an LLM-powered version that offers suggestions interactively. Both versions follow the same risk-envisioning workflow. We evaluated Privy with 24 non-privacy-expert AI practitioners (between-subjects: LLM vs. Template), whose resulting PIA reports were assessed by 13 privacy experts.

Our evaluation study addressed three research questions:

**RQ1** How and to what extent do $\text{PRIVY}_{LLM}$ and $\text{PRIVY}_{Template}$ help practitioners identify and mitigate privacy risks with AI product concepts?

**RQ2** How does the use of large language models affect (i) the quality of privacy impact assessments produced by PRIVY, and (ii) the perceived usefulness of PRIVY in privacy risk envisioning and mitigation?

**RQ3** What challenges do practitioners face in the privacy risk envisioning and mitigation process, and to what extent does PRIVY address these challenges?

Our findings show that PRIVY effectively scaffolds non-privacy-expert practitioners in creating high-quality PIAs. Using both versions of PRIVY, practitioners identified risks that were relevant and severe, and developed effective and appropriate mitigation strategies (RQ1, Section 6.1). Compared to $\text{PRIVY}_{Template}$, $\text{PRIVY}_{LLM}$ produced assessments in which the identified risks were judged by privacy experts to be more relevant, severe, and correct, with clearer articulation. $\text{PRIVY}_{LLM}$ also produced mitigation plans that were more effective and appropriate for the identified risks. Both versions were seen as useful and usable (RQ2, Section 6.2). PRIVY also overcame long-standing barriers that hinder non-privacy-expert practitioners' privacy work by improving awareness (via scaffolding risk identification), fostering motivation (via encouraging reflection and engagement), and supporting ability (by improving privacy self-efficacy) (RQ3, Section 6.3).

In sum, this paper contributes:

- **Five core design goals** for a privacy risk-envisioning tool that helps practitioners identify the most relevant and severe privacy risks in novel AI product concepts, synthesized from a formative study with 11 AI and privacy practitioners (Section 3).
- The **design and implementation of PRIVY**, both $\text{PRIVY}_{LLM}$ and $\text{PRIVY}_{Template}$, that instantiate these design goals (Section 4).
- **Empirical insights** from an evaluation study with 24 non-privacy-expert AI practitioners and 13 privacy experts, demonstrating that PRIVY effectively scaffolds creating quality PIAs (Section 6.1), that LLMs enhance assessment quality through human–AI collaboration (Section 6.2), and that PRIVY helps practitioners overcome awareness, motivation, and ability barriers in privacy work (Sections 6.3).

# 2 Related Work

## 2.1 AI privacy risks

Privacy risks in artificial intelligence (AI) are widely understood as extensions and amplifications of long-standing privacy concerns. Foundational legal frameworks, such as the "Right to Privacy" [69] and tort-based protections against intrusion and appropriation [51], continue to influence how we think about privacy harms. Contemporary scholars have expanded these notions to broader categories of privacy — i.e., informational, behavioral, decisional, and physical [41, 59] — that better reflect the nuanced and multifaceted nature of privacy violations. AI intensifies these risks by extracting insights from vast, often opaque datasets, potentially revealing sensitive information even from anonymized data [61].

Building on legal and ethical foundations, human-computer interaction (HCI) and AI researchers examine how these risks emerge across the AI development lifecycle. For example, Shahriar et al. [56] highlight privacy risks such as re-identification and regulatory non-compliance during early project planning. Martin and Zimmerman [42] further map technical privacy risks to each stage of data collection, processing, and use, highlighting how users perceive the risks and stages. Other work delves into specific AI technologies and associated privacy implications [17]: computer vision technologies heighten concerns around surveillance and physical privacy [46], whereas NLP systems can extract or expose sensitive content from text or speech data [64].

A recurring theme across this literature is that AI does not merely facilitate privacy intrusions, it amplifies them. Recent work in HCI has begun to formalize how AI exacerbates or creates new categories of privacy risks, through updated taxonomies grounded in legal theory and technical capability [40].

Although research has bridged the gap between legal definitions of privacy and AI capabilities, many of the resulting models remain untested by practitioners who develop these systems. There remains a need within the research community for practical artifacts grounded in robust legal and AI privacy scholarship. This paper aims to address this gap by introducing an interactive tool designed to guide AI practitioners through potential privacy challenges identified in the existing literature.

## 2.2 Practitioners' approaches to and challenges with addressing privacy risks

A growing body of research investigates how AI practitioners grapple with privacy risks during the development and deployment of AI systems. Although many practitioners recognize privacy as a critical concern, they often struggle to address privacy risks due to organizational constraints, a lack of supportive policies, and insufficient tools [6, 20, 38]. Pant et al. [48] identify notable challenges in developing ethical and privacy-compliant AI systems: limited training, fragmented understanding of privacy, inadequate foresight, and organizational inertia.

One line of work extends practices from requirements engineering [27, 35] to operationalize privacy by design [5, 31] principles, giving rise to what is now termed "privacy engineering". Within software engineering, this approach conceptualizes two levels at which privacy can be incorporated into a system: *privacy by policy* and *privacy by architecture* [62]. These levels reflect different mechanisms for embedding privacy in product design workflows, be it through user-facing notice and consent controls or through constraining data collection and processing at the product-level.

Yet in practice, privacy work and responsibility are often organizationally siloed. Engineering teams are tasked with executing decisions handed down from legal or compliance teams [30, 36], rather than collaboratively working with these teams to define what privacy should mean for a product and why [38, 44]. This handoff model can introduce inefficiencies and additional rework as teams iterate to align products with regulatory expectations.

These realities underscore the need for tools that support AI development to ensure that privacy becomes a fundamental element rather than an afterthought. Answering this call, we present Privy (Section 4), the first artifact specifically designed to help practitioners, who are not privacy experts, foreground privacy impact assessment for early-phase AI product concepts. Section 3 details our formative study with AI and privacy practitioners to understand their needs for such a tool. Section 5 describes how we tested Privy with real-world AI practitioners to evaluate its impact on their privacy approach in AI product development.

## 2.3 Tooling for privacy risk assessment

Various tools have emerged to help practitioners identify and assess privacy risks. In general, these fall into two categories: **structured worksheets** and **interactive interfaces**.

**Structured worksheets** are static documents that prompt reflection on potential harms. The dominant tool is the Privacy Impact Assessment (PIA) [16], which helps record design decisions and surface privacy risks. These assessments are intentionally technology-agnostic and originate from the need to ensure systematic documentation aligned with legal and regulatory expectations [76]. However, PIAs are frequently regarded as difficult to operationalize, as their guidance on identifying and mitigating privacy risks is often generic and insufficiently tailored to specific technical contexts [4, 26]. These limitations become even more pronounced when PIAs are applied to AI-driven systems, where complexity and opacity challenge the effectiveness of traditional assessment approaches [28]. They rarely provide a comprehensive view of how product data flows might entail both intended and unintended consequences [38]. Nonetheless, ongoing legal and privacy scholarship highlights opportunities to adapt PIAs to be more system-specific and context-aware, thereby improving their relevance and utility in AI-focused environments [49, 67]. PIA worksheets directly informed the privacy-focused approach to Privy, but we contribute to the literature by meeting the calls for AI-specific privacy risks tied to the technology's capabilities and requirements.

Recently developed variations on the worksheets include privacy checklists that guide developers through risk considerations throughout the AI lifecycle [11, 22], and consequence scanning exercises that help teams anticipate both the intended and unintended impacts of AI on users [23]. International standards bodies have also developed frameworks, such as NIST's AI RMF [45], the ISO's Information Technology Security Techniques (29134) [2] and the Artificial Intelligence Set (42001) [3], to promote responsible documentation of development decisions and mitigations. Yet these standards are often static and driven by policy requirements rather than practitioner needs.

**Interactive interfaces** build upon the principles of structured worksheets, but offer digital, dynamic workflows. Some tools integrate stakeholder input with internal privacy expertise to facilitate impact assessments [10], while others use card-based prompts to encourage teams to consider potential effects on diverse stakeholders [24]. Card-based approaches built on the LINDDUN methodology [21] to privacy threat analysis have been shown to lower the barrier of entry for non-privacy-expert practitioners [78]. LINDDUN also offers other methodologies to conduct PIAs for both expert and non-expert privacy practitioners[1]. These tools improve awareness, but still face key limitations: they are slow to adapt, rarely cover the full AI lifecycle (or AI at all), and often lack flexibility in accounting for evolving regulatory or technical contexts [37].

Recent efforts have incorporated generative AI (GenAI) to address these gaps. For example, the Anticipating Harms of AI (AHA) tool helps predict how design decisions can lead to downstream harms [14], while Farsight uses an interactive whiteboard powered by GenAI to generate intended, unintended, and misuse scenarios, each linked to specific harms and real-world case studies [68]. Although these tools aid in harm identification, they offer limited support for ideating mitigations or articulating how specific AI functionalities can lead to privacy risks.

To address these gaps, Privy is designed to guide practitioners through use case ideation, privacy risk identification, and risk mitigation planning, while clearly linking risks to AI system capabilities. Drawing from both structured assessments and interactive tools, we implemented two versions: 1) $\text{Privy}_{Template}$, a *structured worksheet* administered as a Google Doc template, and 2) $\text{Privy}_{LLM}$, an *interactive interface* enhanced with LLM-generated suggestions.

## 2.4 Impacts of GenAI-powered tooling on AI practitioners' privacy work

As GenAI integrates into tools for AI privacy work, researchers highlight both opportunities and trade-offs. Some researchers argue that GenAI supports brainstorming and ideation by offering examples, counterfactuals, and alternative perspectives that improve problem solving [18, 33, 77]. In privacy contexts, this kind of generative ideation can help practitioners envision negative outcomes of proposed use cases, expanding awareness of potential risks.

However, concerns have emerged about practitioners' overreliance on GenAI. Studies show that automation can dampen critical engagement and contribute to workforce deskilling, especially among junior practitioners [32, 53, 75]. These risks are particularly salient in AI privacy, where practitioners' expertise is inconsistent and potentially underdeveloped [38, 48].

Nonetheless, GenAI can assist with privacy work by issuing early warnings of oversights. Emerging research explores GenAI as a defense mechanism in cybersecurity and privacy, helping reduce the cognitive and operational burden on human operators [29, 57]. When implemented carefully, GenAI has the potential to serve as a valuable partner in AI privacy work.

To investigate these possibilities, we explore the potential impacts of GenAI-powered tooling on AI privacy practices (RQ2). In particular, we examine how practitioners use these tools, the benefits they observe, and the challenges they face.

# 3 Formative Study

Since there is no established standard for AI privacy impact assessments (Section 2.3), we explored the design opportunities for this new tool through formative design engagements with an early prototype, $\text{Privy}_{beta}$. We first created an interface for $\text{Privy}_{beta}$ that adapted the "use cases → stakeholders → harms workflow from Farsight [68] — a general, LLM-powered AI risk-envisioning system — to the context of AI privacy impact assessments. Specifically,

[1]https://linddun.org/

Privy$_{beta}$ included two core LLM-powered interactive interfaces for practitioners to generate AI privacy impact assessments for AI product concepts: (1) the *risk explorer*, which guided users through articulating a product concept's capability and requirement, use cases, and impacted stakeholders to identify the most pertinent of 12 types of AI privacy risks modeled from the privacy taxonomy [40]; and (2) the *risk summarizer*, which synthesized users' identified risks and surfaced related real-world incidents sourced from the AIAAIC, a popular AI incident database [50]. The two interfaces can be found in Appendix C.1.

We used this early prototype to inform Privy's design by conducting a formative study with $N = 11$ practitioners. We asked: *how can we design Privy to better support practitioners in identifying AI privacy risks and making privacy-informed design decisions for novel product concepts?* From this study, we derived five design goals (DGs) that guided the final design of Privy (DG1–DG5; Section 4).

## 3.1 Participants

We recruited 11 AI and privacy practitioners (nine in the US, two in the UK) spanning eight companies through the authors' direct contacts and social networks. We required participants to have experience with designing and/or developing consumer-facing AI products at work. Each 60-minute session was compensated with a $100 Amazon gift card. Participants included eight men and three women with diverse roles (Appendix C.2). Nine reported having engaged in end-user privacy discussions as part of their work.

## 3.2 Procedure

Each session began with participants describing their experience identifying privacy risks for an early-stage consumer-facing AI product they had worked on. After being introduced to Privy$_{beta}$, they used the *risk explorer* to identify risks in that AI product and then reviewed a one-page privacy report in the *risk summarizer*. Participants were asked to think aloud during the session and provide feedback on Privy$_{beta}$'s features. We concluded with a semi-structured interview (Appendix C.3) where participants reflected on Privy$_{beta}$'s potential fit in their product development and suggested alternative design ideas.

All study sessions were audio- and video-recorded and then transcribed. Two researchers applied iterative open coding [55], in which they coded the transcripts independently, and met regularly to resolve any conflicts in coding and identify emerging use patterns and themes.

## 3.3 Findings

*3.3.1 Scaffold product-specific AI privacy implications.* Practitioners reported that Privy$_{beta}$ guided them to systematically identify privacy risks in their AI product concepts. It foregrounded aspects they might otherwise overlook — such as unintended use cases, stakeholders, and data requirements. U8, who worked on a customer-facing AI agent, reflected: *"This is helpful for seeing what types of data... initially [I] didn't really think about the capabilities, but the underlying requirements and how that really requires sensitive user data."* Similarly, the tool's unintended use cases surfaced misuses practitioners were hesitant to assume (*"don't really like assuming our user doing something pretty bad"*, U6) or risks they had not considered beyond compliance (*"only aware [of privacy risks] when it comes to... compliance of the laws"*, U9).

Overall, practitioners benefited from **a structured workflow that links AI product concepts to privacy risks and valued structured support to guide them through it**.

*3.3.2 Facilitate privacy discussion and decision-making.* Practitioners found the risk summarizer helpful for initiating privacy discussions and driving decisions. The interface condensed product-specific risks into key insights, which U9 described as *"taking the most important messages from the exploring page [risk explorer] to this summary."*

While prior work shows privacy is often treated as an afterthought for compliance [52], participants appreciated how Privy$_{beta}$ enabled a more proactive review of AI product concepts, and extrapolated its impacts on their product development. U3 reflected: *"usually customers provide the requirement and we fulfill it, but now we can... initiate discussion with them based on potential privacy issues."* Others emphasized its alignment with the aspirational *'Privacy by Design'* ideal, catching risks early and reducing downstream rework. As U6 explained: *"That will save us a lot of time... we didn't think about the privacy thing until the end... and then we had to spend another two weeks to work with other engineers."*

Overall, practitioners valued having **an artifact that supports proactive privacy discussions and informed decision-making**.

*3.3.3 Enhance AI privacy risk awareness.* Privy$_{beta}$ aimed to raise practitioners' awareness of AI privacy risks through product-specific vignettes illustrating how risks might manifest. Participants wanted these vignettes to be more explicitly tied to their products. They sought a clear line from the capabilities and requirements of their AI concepts to the identified risks. As U10 explained: *"It would be helpful to have some of these risks directly tied to specific capabilities or features... You need to map these [risks] back to the original functionality."*

Participants also noted that the volume of risks could be overwhelming, especially for non-privacy experts. Practitioners expressed a need for prioritization to better triage in resource-constrained environments. As U5 put it: *"The most important thing for me is to see the information, but also to have the tool help me prioritize all the risks so I can decide what to focus on next."*

Overall, practitioners needed **clear connections between product capabilities and privacy risks**, and **support for prioritizing the most important risks**.

*3.3.4 Emergent ideas: support mitigation and autonomy in privacy work.* Our study also surfaced new directions for Privy. Practitioners often moved from identifying risks to considering mitigation, and wanted actionable next steps. As U5 put it: *"It would be really nice to have some sort of AI-powered actionable item to suggest for next steps. I don't have to follow them, but at least I have better ideas on where to start."*

Participants also sought greater control over Privy$_{beta}$-generated content to ensure it aligned with their work. Some modified or expanded Privy$_{beta}$'s outputs to better capture product nuances, while others critiqued the outputs for falling short of their standards for work (e.g., novelty, creativity): *"it didn't think outside the box as much as like a professional would"* (U10).

Overall, practitioners wanted **support for developing risk mitigation strategies**, and **more precise control over LLM-generated content.**

# 4 Privy

From our formative study findings (Section 3.3), we distilled five design goals (DG1–DG5) to guide the creation of Privy, a tool that supports AI practitioners in envisioning and mitigating privacy risks for novel AI product concepts. Privy leads users through a structured workflow: they first enter a natural language description of the product and its data sources; then outline use cases and associated beneficiaries; specify the AI capabilities and requirements to support those use-cases; and, finally, identify the most relevant and severe risks from the AI privacy taxonomy [40] based on these capabilities and requirements, use-cases, and the product description. We implemented two versions of Privy, modeled after prior AI risk assessment approaches (Section 2.3): (1) $\text{Privy}_{LLM}$, an interactive system that leverages LLMs for human–AI collaboration, and (2) $\text{Privy}_{Template}$, a structured worksheet. Both embody the same scaffolded workflow. In the following sections, we use $\text{Privy}_{LLM}$'s interfaces to illustrate how we achieve each design goal, and highlight LLM-powered features that support human–AI collaboration. We present the $\text{Privy}_{Template}$'s interface in Section 4.5.

## 4.1 DG1: Help practitioners elicit the AI capabilities and requirements that entail privacy risks

Our formative study showed that practitioners' ability to identify privacy risks is grounded in understanding *what their AI product does and what it requires to work.* Likewise, prior work finds that practitioners feel more accountable for harms when reasoning about how *their own* products may cause them, rather than abstracting harms [71]. To that end, Privy helps practitioners articulate capabilities and requirements of their AI product concepts grounded in their envisioned use cases (Section 4.1.1); $\text{Privy}_{LLM}$ augments this process with LLM-powered suggestions (Section 4.1.2).

*4.1.1 AI Capability & Requirement Scaffolder.* Both $\text{Privy}_{Template}$ and $\text{Privy}_{LLM}$ guide users through three steps: (1) describing their AI product concepts, (2) brainstorming use cases, and (3) summarizing resulting AI capabilities and requirements.

*Product description node.* Users first draft a description of their AI product concept — its purpose, required data, and an example use case (Fig. 1A). For example: *AI Meeting Assistant* may be described as *"a feature that summarizes meetings and follow-ups"*, requiring *"access to audio/video streams,"* with a use case of *"deriving action items from a meeting."*

*Use case node.* Next, users brainstorm *intended* and *unintended* use cases (Fig. 1B-1). Privy also prompts them to specify *beneficiaries* of intended uses and *exploiters* of potential misuses. For example, one use case might be: *"a user employs AI Meeting Assistant to generate action items"* with *"meeting participants"* as the beneficiaries.

*AI capabilities and requirements summary node.* Privy guides users to distill the AI capabilities and corresponding requirements of their product concept based on envisioned use cases (Fig. 1C). For example, an AI Meeting Assistant might be described as having the capability to *"recognize key action items"*, which requires *"access to calendar data and device microphones."*

*4.1.2 Human-AI Collaboration in Scaffolding AI Capabilities and Requirements.* For $\text{Privy}_{LLM}$, we leverage an LLM to help users brainstorm more use cases and summarize their AI products' capabilities and requirements (see Supplementary Material for LLM prompts). For this paper, we specifically used OpenAI's GPT-4.1 model, though Privy could work with any back-end LLM that supports the Chat Completions API.

*AI-assisted use case brainstorming.* From the product description, the LLM suggests intended and unintended use cases with the corresponding beneficiaries or exploiters. Each use case is shown as a sentence describing how a user could use the AI product in a specific context — e.g., *"a user can use a textbook tutor AI to generate a list of recommended readings based on prior coursework"*. To diversify generated use cases, we instruct the LLM to consider both *high-stakes* (e.g., *college entrance exams*) and *low-stakes* (e.g., *take-home quizzes*) contexts. $\text{Privy}_{LLM}$ initially offers four examples (intended/unintended × high-/low-stakes) and allows users to request more (Fig. 1B-2).

*AI-assisted capability and requirement summarization.* We prompt LLMs to extract an AI product's capabilities using the AI capabilities taxonomy [79], and to infer associated requirements based on data collection, processing, dissemination, and infrastructure needs [40]. First, the LLM generates capability–requirement pairs from each envisioned use case. For example, from the use case *"a user requests recommended readings based on prior coursework"*, $\text{Privy}_{LLM}$ derives the capability to *"generate personalized reading recommendations and factual responses by analyzing textbook content and students' historical coursework"*, along with the following requirements: *"(1) collect textbook materials and student learning histories, (2) process and analyze these data to generate recommendations, (3) disseminate recommendations to users, and (4) integrate with learning platforms for user interaction and data exchange"*. A second LLM then merges all pairs into a concise summary, shown in $\text{Privy}_{LLM}$'s summary node (Fig. 1C).

## 4.2 DG2: Help practitioners identify and prioritize AI privacy risks

Practitioners lack tools to effectively identify and prioritize AI privacy risks [38]. Privy addresses this gap by guiding users to envision, assess, and rank risks drawn from an AI privacy taxonomy [40].

*4.2.1 Privacy Risk Explorer.* Privy's *privacy risk assessment node* enables users to select risks that are relevant and severe, describe how their AI product may elicit those risks, and prioritize them for mitigation. It is complemented with two key components: a *risk definitions panel* and a *risk ranking panel.*

*Risk definitions panel.* This panel presents definitions of each risk type from the AI privacy taxonomy [40], with external links

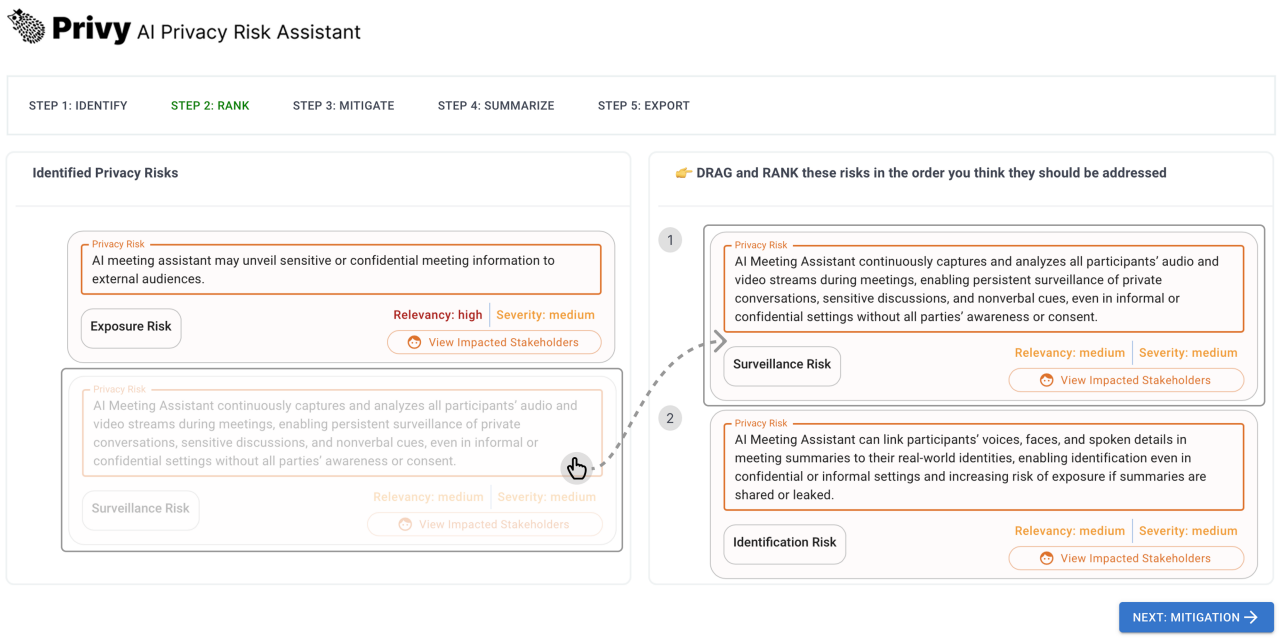

**Figure 2: *Risk ranking panel:* users rank the identified risks in the order they think they should be addressed via a drag-and-drop interface.**

to real-world incidents sourced from the AIAAIC database [50] (Fig. 1E).

*Risk assessment and ranking.* In the privacy risk assessment node (Fig. 1D-1), users describe how each selected risk may arise in their product and who may be impacted. They rate each risk's *relevancy* and *severity* (High, Medium, Low). Then, users drag and drop risks into a ranked list via the *risk ranking panel* (Fig. 2).

#### 4.2.2 Human-AI Collaboration in Envisioning Privacy Risks.

*AI-assisted privacy risk envisioning.* We use a few-shot prompting approach [13] to generate examples of privacy risks for users' AI product concepts, with the associated stakeholders, tailored to the product's description, use cases, and AI capabilities and requirements. For example, for a textbook tutor AI, Privy$_{LLM}$ may suggest a *surveillance risk* such as *"monitoring individuals' learning histories and performance"*, particularly for *"students who use the platform"*. Privy$_{LLM}$ produces one example per privacy risk type defined in Lee et al. [40] (12 in total), which users can review and selectively incorporate into their assessments (Fig. 1D-2).

## 4.3 DG3: Help practitioners brainstorm potential mitigation strategies for AI privacy risks

The ultimate goal of identifying privacy risks is to *address* them. In our formative study, we found that practitioners consider risk mitigation even when identifying risks. Moreover, fully addressing privacy risks can be an iterative and collaborative process, requiring the joint effort of the entire product team. To that end, brainstorming privacy risk mitigation strategies in Privy is intentionally open-ended: outputs from Privy are meant to be a conversation starter rather than the final mitigation strategy.

#### 4.3.1 Privacy Risk Mitigator.

Privy provides an interface that allows practitioners to brainstorm and integrate mitigation ideas across risks. Privy prompts users to outline a mitigation plan for one identified risk at a time, while the mitigation plan is also carried over to the next risk. In this way, users can iteratively build their mitigation strategies, as one part of the strategy may address multiple risks. Users are also asked to rate their confidence in their mitigation plan for each risk they identified. This encourages further reflection on the viability of their plans (Fig. 3).

#### 4.3.2 Human-AI Collaboration in Brainstorming Mitigation Strategies.

*AI-assisted risk mitigation brainstorming provocations.* To support ideation, Privy$_{LLM}$ generates open-ended questions tailored to users' AI product concepts, encouraging users to consider how their systems might incorporate privacy-preserving features or practices in response to specific risks. For instance, if a textbook tutor AI raises *surveillance risks* due to monitoring students' learning history, a mitigation prompt might encourage reflection on *"how to provide users with tools to control, limit, or delete collected data"*.

We grounded these provocations in the Security and Privacy Acceptance Framework (SPAF) [19], which identifies three key barriers to end-user privacy adoption: lack of *awareness*, *motivation*, and *ability* to address the risk. To scaffold product-specific reflection, the LLM is instructed to first propose three product-specific features or practices — each targeting one of the SPAF barriers — and then "flips" them into open-ended brainstorming questions that users can use to arrive at appropriate mitigation strategies. Users can request these tailored provocations on demand (Fig. 3A).

## 4.4 DG4: Generate an AI privacy impact assessment report that practitioners can share and use

Products are developed in teams, and practitioners create privacy impact assessments to shape how their entire team approaches privacy risk mitigation. Privy supports this goal by enabling practitioners to easily compile and share findings.

#### 4.4.1 Privacy Risk Summarizer.

Privy collates outputs from the capability & requirement scaffolder, risk explorer, and risk mitigator into a shareable privacy impact assessment report. The interface is modeled after an established AI impact assessment report template [11] and presents three sections (Appendix A): (1) product description, use cases, and AI capabilities & requirements; (2) identified privacy risks along with relevancy and severity ratings provided by practitioners; and (3) proposed mitigation strategies.

Each report is generated automatically and can be shared with collaborators via a unique link.

## 4.5 DG5: Ensure autonomy and control in practitioners' privacy work

Privy is designed to empower non-privacy-expert practitioners to independently identify and mitigate AI privacy risks. To support this, Privy encourages user reflection and ownership over the privacy assessment process through editable interfaces and intentional design friction.

#### 4.5.1 User Agency in Human-AI Collaboration.

*Editable nodes and report interface.* All content in the privacy impact assessment report is reviewable and editable throughout the workflow — including entries from the capability & requirement scaffolder, risk explorer, mitigator, and summarizer (e.g., Fig. 4).

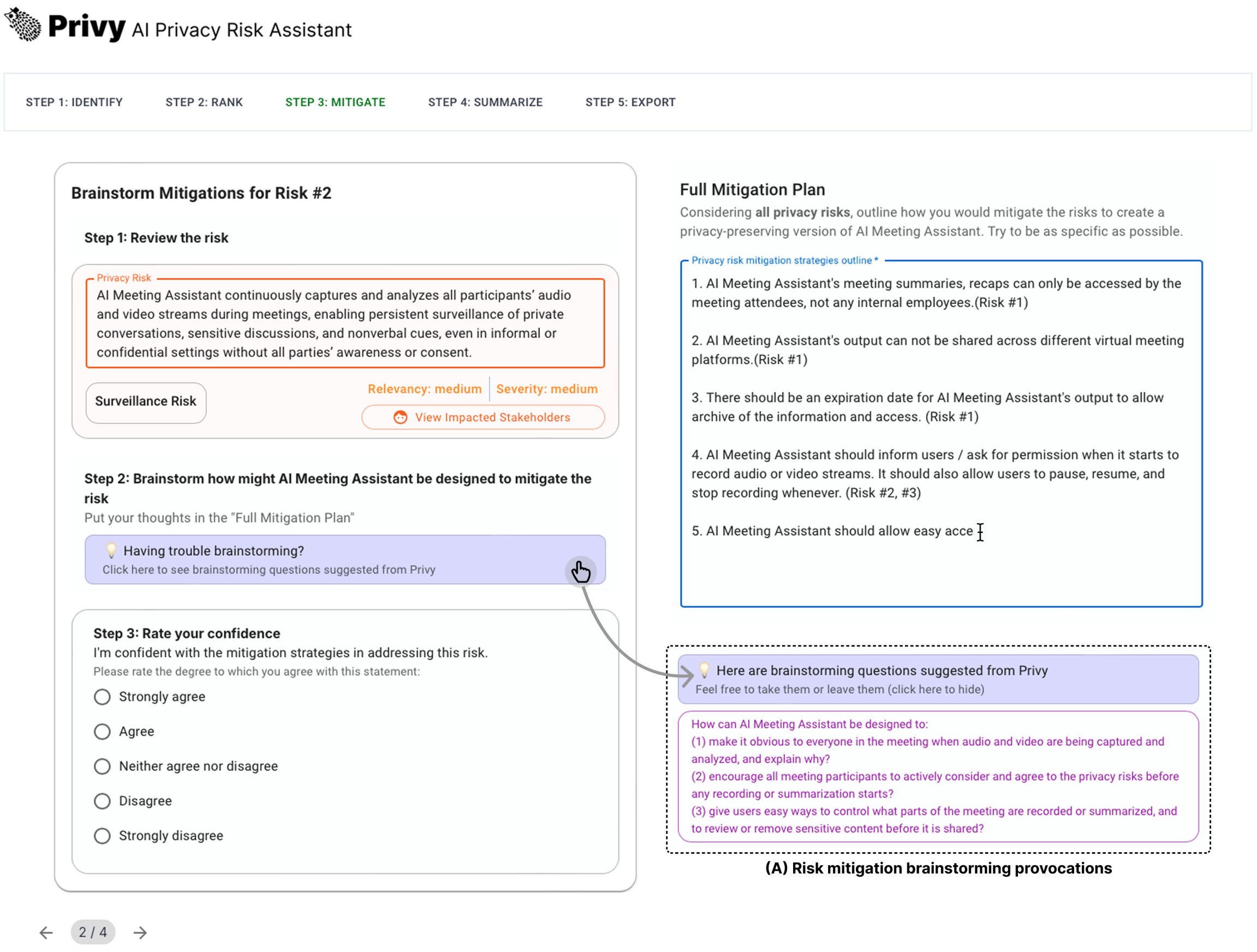

**Figure 3: Privacy Risk Mitigator: users brainstorm a risk mitigation plan for risks they identified, and receive (A) *risk mitigation brainstorming provocations* to assist their brainstorming.**

Users retain full control over how their assessment is constructed and shared.

*Design friction.* To promote *critical integration* of LLM outputs [39], we embed design friction into Privy$_{LLM}$. For example, users must manually assess the relevance and severity of AI-suggested risks before including them (Section 4.2.1). Similarly, when brainstorming mitigations, users receive provocative questions rather than concrete suggestions — encouraging them to generate context-appropriate solutions themselves (Section 4.3.2).

*4.5.2 Privy$_{Template}$: Privy as a structured worksheet.* To isolate the role of LLM assistance, we developed Privy$_{Template}$, a static version of Privy that follows the same workflow (DG1–DG4) without LLM assistance. Delivered as a structured Google Doc, it includes the full AI privacy taxonomy [40] (Fig. 5A) and guides users through three sections: Product Information, Privacy Risks, and Mitigation Strategies (Fig. 5; full version in Appendix B).

For mitigation, Privy$_{Template}$ substitutes dynamic LLM-generated provocations with static ones grounded in the SPAF [19]: *How can the product be designed to 1) enhance users' **awareness** of the risk? 2) enhance users' **motivation** to address the risk? 3) provide users with the **ability** to manage the risk?* Privy$_{Template}$ builds on an established AI impact assessment report template [11], ensuring usability and shareability (DG4).

In our evaluation, comparing Privy$_{LLM}$ and Privy$_{Template}$ allowed us to assess how LLM assistance affected the quality of privacy assessments and practitioners' perceived usefulness of Privy.

## 4.6 Privy$_{LLM}$ Implementation

We implement Privy$_{LLM}$ as a web client: we implement the interactive front-end with React.js, and use a Python Flask server as the backend. We leverage React Flow[2] and Material UI[3] to implement node-link interactions and styling. We leverage a large language model, GPT-4.1, to enable the human-AI collaboration features we described, and follow best practices when designing our prompts, including the use of few-shot learning and chain-of-thought prompting [70]. For the few-shot examples, we manually curated 23 different AI product concepts across ten types of AI technology [4], referenced on real-world AI products sourced from the AIAAIC database [50]. We included our LLM prompts, few-shot examples, and additional LLM-generated examples (e.g., use cases, privacy risks) in the Supplementary Materials.

[2]https://github.com/xyflow/xyflow/tree/v11
[3]https://github.com/mui/material-ui

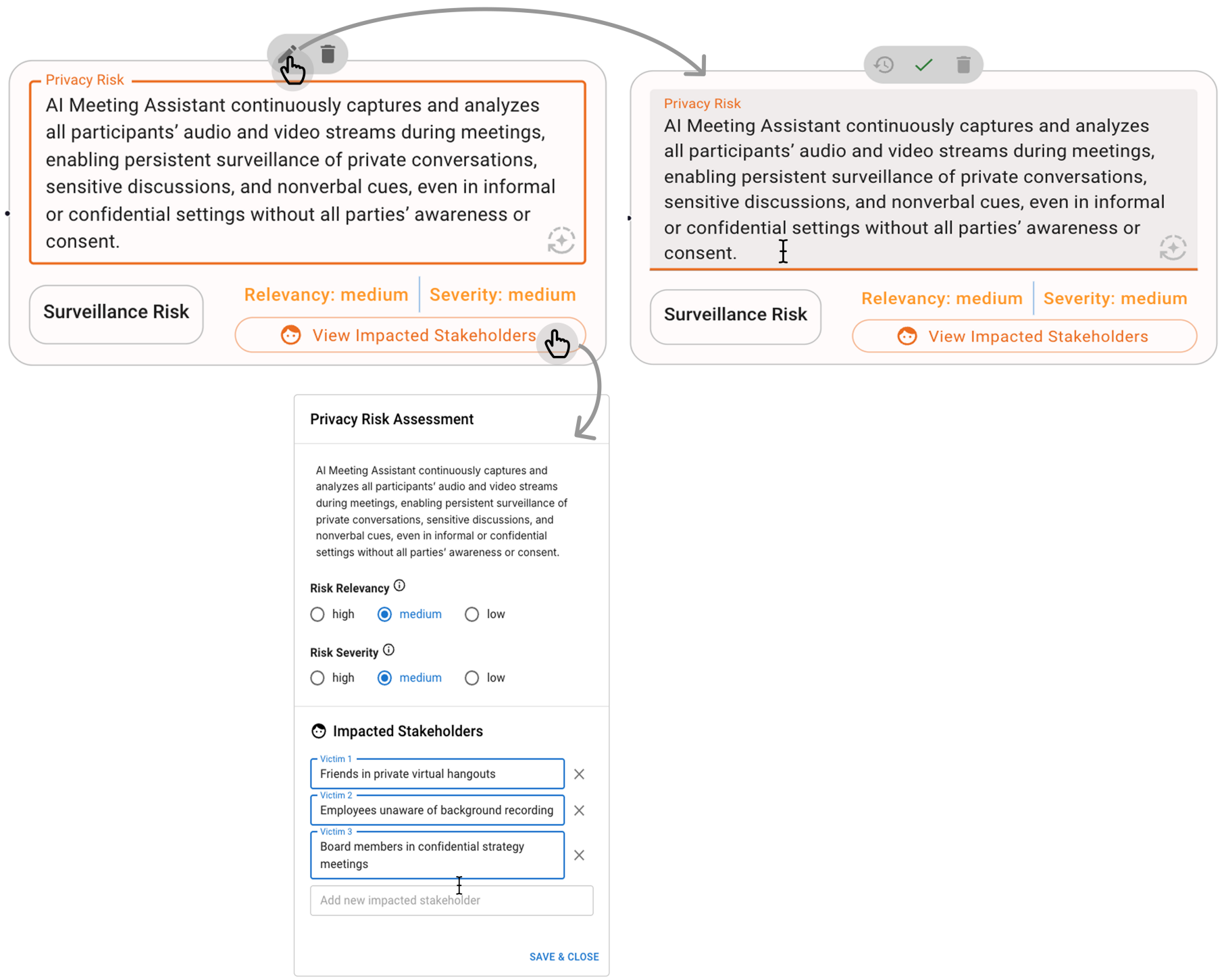

**Figure 4:** ***Privacy risk assessment panel:*** **users can edit the node content to capture their own perspectives, including the risk description, impacted stakeholders, and their assessment of the identified risk's *relevancy* and *severity*.**

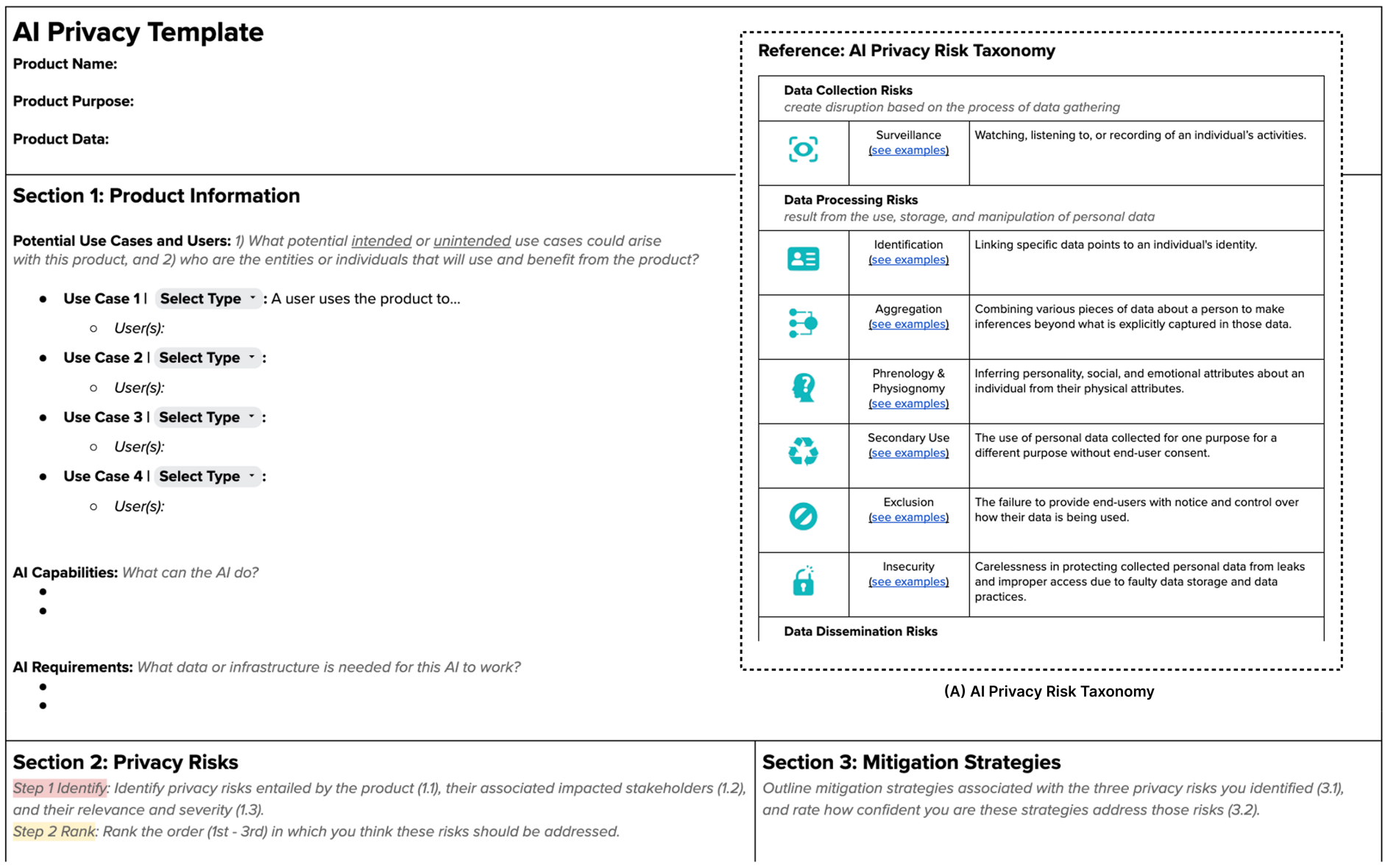

**Figure 5: Privy$_{Template}$: the worksheet is administered in the form of a structured Google Doc with three main sections: Product Information, Privacy Risks, and Mitigation Strategies. The worksheet contains (A) the *AI privacy taxonomy* [40] in a detailed table as a reference. See the full worksheet in Appendix B.**

# 5 Evaluation User Study

To address our three research questions (Section 1), we conducted a between-subjects study with $N = 24$ industry AI practitioners who were not privacy experts but worked across diverse AI product roles. Participants were randomly assigned to use either *PRIVY*$_{LLM}$ ($N_{llm} = 12$) or *PRIVY*$_{Template}$ ($N_{template} = 12$) to create privacy impact assessments for two pre-defined AI product concepts. We also recruited $M = 13$ privacy experts — who work as privacy practitioners and/or researchers in industry or academia — to independently evaluate the resulting reports; these experts were not told which reports were produced by PRIVY$_{LLM}$ or PRIVY$_{Template}$. Qualitatively (RQ1–RQ3), we analyzed participants' think-aloud and interview transcripts. Quantitatively (RQ2), we compared practitioners' self-report perceptions and expert-rated report quality across conditions (PRIVY$_{LLM}$ vs. PRIVY$_{Template}$).

## 5.1 Participants

We recruited 24 industry practitioners with experience designing or developing consumer-facing AI products, drawing from the research team's professional networks. Twenty participants were based in the US and four in the UK. To avoid internal validity bias, subjects from the formative study were barred from participating.

We conducted four pilot studies with PRIVY$_{LLM}$ and four with PRIVY$_{Template}$ (eight total). Each study session lasted 90 minutes, and participants received a $150 Amazon gift card as compensation. Of the 24 participants, 17 identified as men and seven as women, spanning a range of industry roles (Appendix D.1). None were privacy experts: four reported being knowledgeable about privacy, 15 had some knowledge, four had limited knowledge, and one had none.

## 5.2 Study Design

We conducted one-on-one remote sessions over Zoom, recording video, computer screens, and audio for analysis. Participants were randomly assigned to one of four conditions: using either *PRIVY*$_{LLM}$ or *PRIVY*$_{Template}$ to assess one of two pre-defined AI product concepts (Table 1), with six participants per condition.

Our between-subjects design examined how LLM-assistance shaped (i) participants' approach to risk identification and mitigation, (ii) the quality of privacy impact assessments practitioners produced, and (iii) practitioners' subjective perceptions towards the process (RQ2). To strengthen this evaluation, we combined two perspectives: (1) practitioner data from think-alouds, surveys, and interviews, and (2) expert evaluations from 13 privacy experts who independently assessed the quality of the resulting privacy impact assessment reports (Section 5.2.3). **Using two product concepts also helped generalize our findings beyond a single product type.**

[4]The ten types of AI technologies are: computer vision, generative AI, speech-voice system, decision support system, information retrieval, natural language processing, user modeling, predictive analytics, recommendation system, and conversational AI.

Each 90-minute session comprised: (1) a privacy impact assessment activity (Section 5.2.1) where participants identified three privacy risks and proposed mitigations using their assigned PRIVY version (RQ1), and (2) a post-activity interview (Section 5.2.2) to explore participants' tool feedback and experiences more broadly (RQ3). The study was approved by our institutional review board (IRB).

### 5.2.1 Privacy Impact Assessment (PIA) Activity.

For each PIA activity, which included both risk identification and risk mitigation, participants were presented with a description of an AI product concept, including its purpose and the data it required (Table 2). We chose these two AI product concepts that are representative of two common types of consumer-facing AI technologies (i.e., recommendation systems, AI-powered assistants) that have been used as hypothetical product examples in prior risk assessment studies [11, 25].

During risk identification, participants described three risks — from the twelve total in the AI privacy taxonomy [40] — that they felt the most important to mitigate. Next, during risk mitigation, they proposed alternative designs of the AI product concept or other strategies (e.g., legal/policy reviews, data access controls) that they would like to share with their team to mitigate the three identified risks. Participants were instructed to work entirely within PRIVY's features and interfaces, and to think-aloud throughout. Participants were also asked to reflect on their approach, once after completing risk identification and once after completing risk mitigation. Finally, we reviewed the final report — including the identified risks and proposed mitigation plan — together with the participants for any final adjustments before proceeding to the post-activity interview. We include the privacy risks identified and the proposed mitigation plans from each participant in the Supplementary Materials.

*PIA Procedure.* Each session began with participants rating the assigned AI product concept's perceived *benefit* (e.g., "AI Meeting Assistant is beneficial to its envisioned users") and *intrusiveness* (e.g., "AI Meeting Assistant is risky to the privacy of its envisioned users, other stakeholders, or society at large.") on a 5-point Likert scale. After a brief demo of the assigned tool, they completed the PIA activity (25 minutes for identification, 15 minutes for mitigation).

### 5.2.2 Post-activity Semi-structured Interviews.

Following the PIA activity, participants re-rated the AI product concept's benefit and intrusiveness, and completed survey items (Appendix D.2.2) on the tool's usefulness (PRIVY design goal alignment) and usability (System Usability Scale [12]). Semi-structured interviews (Appendix D.2.1) probed how PRIVY influenced their approach to privacy, tool feedback, and broader challenges in conducting PIAs.

### 5.2.3 Privacy Impacts Assessment Report Evaluation by Privacy Experts.

We collected a total of 24 PIA reports comprising 72 identified privacy risks (with product-specific descriptions of those risks) and 24 mitigation plans. To evaluate the quality of these reports, we recruited 13 external privacy experts (ten industry, three academia) through our networks. All self-identified as privacy experts, with roles spanning researchers, privacy engineers, privacy program managers, consultants, and a professor. None had prior exposure to PRIVY.

**Table 1: Each participant was randomly assigned to one of the four study conditions ( 2 Privy versions x 2 AI product concepts, six participants per condition).**

| Condition (Privy version & AI product concept) | Participant IDs |
|---|---|
| $\text{Privy}_{LLM}$ & AI Meeting Assistant | P1, P2, P3, P4, P5, P6 |
| $\text{Privy}_{LLM}$ & Dynamic Audience Selection | P7, P8, P9, P10, P11, P12 |
| $\text{Privy}_{Template}$ & AI Meeting Assistant | P13, P14, P15, P16, P17, P18 |
| $\text{Privy}_{Template}$ & Dynamic Audience Selection | P19, P20, P21, P22, P23, P24 |

**Table 2: Descriptions of the two AI product concepts used in the evaluation study: AI Meeting Assistant (left) and Dynamic Audience Selection (right).**

| AI Meeting Assistant | Dynamic Audience Selection |
|---|---|
| *Product Purpose:* An online meeting software feature that helps users summarize meeting notes, highlights, and key follow-up items. The feature works in the background during an online meeting and sends the meeting summary to the users once the meeting is completed.<br><br>*Product Data:* The feature needs access to users' video and audio data stream via the cameras and microphones on their devices. | *Product Purpose:* A feature on a social media platform that helps users identify user groups and allow them to dynamically specify audience groups of a post they would like to include with natural language using "+" markers, and those they would like to exclude using "-" markers, e.g., "+friends in music groups, -relatives".<br><br>*Product Data:* The feature needs access to users' posts and contact information on the social media platform. |

Each expert reviewed four PIA reports for the same product concept (AI Meeting Assistant or Dynamic Audience Selection), along with the accompanying product description (Table 2). We originally aimed for 12 experts, each evaluating four reports (two from $\text{Privy}_{LLM}$, two from $\text{Privy}_{Template}$), such that every report was evaluated twice. Through our parallelized recruitment outreach, we ultimately recruited an additional privacy expert, resulting in four reports being evaluated by three experts[5]. Experts were blinded to the study conditions ($\text{Privy}_{LLM}$ vs. $\text{Privy}_{Template}$), and the reports they reviewed were randomly and evenly sampled.

The identified risks and the mitigation plan of the same PIA report were evaluated together to provide a complete context. For each report, experts first rated three risks (*risk type* selected from the AI privacy taxonomy + *description* of how the risk might be manifested) on *relevancy* and *severity*, the items adapted from the common likelihood-severity risk assessment framework [11, 68]. They were asked to evaluate the quality of the content of these risk descriptions with respect to their *correctness* and *clarity*.

Each mitigation plan included a list of brainstormed strategies to mitigate the identified risks, with indices of how each strategy aimed to resolve which specific risk(s). Mitigation plans were rated on *effectiveness* and *appropriateness*. We defined effectiveness as whether a plan would *address each of the identified risks*, as well as whether it would be a *useful conversation starter* for developing a more comprehensive plan. We defined appropriateness as whether the content of a mitigation plan was *specific* to the product concept and *practical*. Finally, experts rated whether they could envision producing a similar PIA report themselves, used as a proxy for overall quality (Table 3). All items used a 6-point Likert scale (strongly disagree, disagree, slightly disagree, slightly agree, agree, strongly agree). During data analysis, we set these six categories as ordinal scores: 1, 2, 3, 4, 5, and 6.

## 5.3 Data Analysis

We applied a mixed-methods approach to analyzing our data. **Quantitative analysis.** We calculated descriptive statistics (means, standard deviations) to examine how $\text{Privy}_{LLM}$ and $\text{Privy}_{Template}$ influenced practitioners' ability to identify and mitigate privacy risks (RQ1). We fit ordinal logistic regressions to test how Privy version affected (i) privacy experts' quality ratings of reports and (ii) practitioners' self-reported usefulness of Privy in risk envisioning and mitigation (RQ2). Privy version was the main predictor ($\text{Privy}_{Template}$ as reference). For expert evaluations, we included random intercepts for expert ID and, where applicable, participant ID to account for repeated measures.[6]

**Qualitative analysis.** We analyzed screen recordings and transcripts, including think-alouds and post-activity interviews. Following our research questions, we applied iterative open coding [55]. Three authors performed the initial coding on six interview transcripts to iteratively develop a codebook, then individually coded the remaining transcripts, meeting regularly to refine codes and resolve conflicts. In the following section, we report themes organized by research question and include the final codebook in Appendix D.3.

# 6 Evaluation User Study Findings

## 6.1 RQ1: How and to what extent do $\text{Privy}_{LLM}$ and $\text{Privy}_{Template}$ help practitioners identify and mitigate privacy risks with AI product concepts?

We first analyzed how Privy, which in both versions helps foreground privacy considerations in designing AI products, shapes

[5]For the additional privacy expert evaluation (E13), we used the same procedure to randomly sample four assessment reports (two from $\text{Privy}_{LLM}$, two from $\text{Privy}_{Template}$) of the AI Meeting Assistant product concept.

[6]Specifically, for the "Per Risk" quality measure in Table 3, we included random intercepts for both expert ID and participant ID.

**Table 3: Quality measures and the questions used in the privacy expert evaluation.**

| Evaluation Unit | Quality Measure | Question |
|---|---|---|
| Per Risk | *Relevancy* | The "risk description" is relevant to the product as defined by its purpose. |
| | *Severity* | The "risk description" could significantly impact society and/or specific stakeholder groups if realized. |
| | *Correctness* | The "risk description" aligns with the "risk type". |
| | *Clarity* | The "risk description" is clear. |
| | *Addressing Identified Risks* | The "risk mitigation plan" is effective in addressing this risk. |
| Per Plan | *Useful Conversation Starter* | The "risk mitigation plan" is a good starting point for the product team to address the privacy risks entailed by the product idea. |
| | *Product Specificity* | The "risk mitigation plan" is tailored to the product. |
| | *Practicality* | The "risk mitigation plan" is practical. |
| | *Overall Risk Envisioning* | I could see myself giving a similar privacy risk assessment (i.e., risks described) for this product idea. |
| | *Overall Risk Mitigation* | Based on the identified risks, I could see myself coming up with a similar risk mitigation plan. |
| | *Overall Risk Impact Assessment* | Taking both the "identified risks" and the "risk mitigation plan" into account, this is a high quality privacy risk assessment. |

**Table 4: Privacy expert evaluation on the identified privacy risks' relevancy and severity.**

| Measurement: 6-point Likert Scale* | PRIVY (Overall) | PRIVY$_{LLM}$ | PRIVY$_{Template}$ |
|---|---|---|---|
| **Relevancy** | 4.58 (SD=1.32) | 5.09 (SD=1.05) | 4.06 (SD=1.36) |
| **Severity** | 4.97 (SD=1.07) | 5.46 (SD=0.68) | 4.49 (SD=1.17) |

*Note.* 1: strongly disagree, 2: disagree, 3: slightly disagree, 4: slightly agree, 5: agree, 6: strongly agree

practitioners' approaches to risk identification (Section 6.1.1) and risk mitigation (Section 6.1.2). Since the findings reported in this section are shared across both PRIVY$_{LLM}$ and PRIVY$_{Template}$, we refer to them collectively as "PRIVY" for simplicity.

*6.1.1 PRIVY helps practitioners envision privacy risks that are relevant and severe.* We found that PRIVY supported participants in identifying a broad spectrum of privacy risks for each AI product concept — practitioners surfaced at least eight different types of privacy risks, with each of those risk types being raised by two or more participants (Appendix D.4). Our findings suggest that PRIVY helped participants identify a diverse set of privacy risks that were both severe and relevant to their AI product concepts (DG1, DG2), from both the practitioners' own and the privacy experts' perspectives. Specifically, participants rated 55 (76%) of the 72 privacy risks they themselves identified as being of high relevancy and 47 (65%) of high severity. Similarly, on a 6-point Likert scale (6 as strongly agree), privacy experts, on average, rated these risks as 4.58 (SD=1.32) on relevancy and 4.97 (SD=1.07) on severity (Table 4).

Our qualitative findings illustrate how PRIVY supported participants beyond relying only on their own privacy expertise. Guided by the *AI Capability & Requirement Scaffolder* (Section 4.1.1) and *Privacy Risk Explorer* (Section 4.2.1), participants grounded envisioned risks in product-specific **use cases**, **AI capabilities**, and **AI requirements**.

*Grounding risk identification with use cases.* Practitioners often drew on the use cases they generated, along with envisioned beneficiaries and exploiters for those use cases, when articulating privacy risks. As P18 noted about AI Meeting Assistant: *"managers can misuse this tool... to inaccurately assess webinar... the second [use case] that I mentioned here is... checking the portion of the meeting participant dialog to assess how people engaged in the meetings."* Use cases also helped participants surface risks for indirect stakeholders. For example, P19 identified that Dynamic Audience Selection could reveal personal information of users' contacts on social media: *"it's less impact on the stakeholders or the user, but more about those who you share with [AI], in your contact form or the profile."*

*Grounding risk identification with envisioned AI capabilities.* Participants connected risks to the AI capabilities they brainstormed. For instance, P21 noted that Dynamic Audience Selection could *"infer the [users'] personality based on how they interact with posts... and then compare that with their physical image."* Participants also anticipated risks when such capabilities fail or underperform: *"if I said, 'minus relatives,' but then it [Dynamic Audience Selection] shared it with some relatives anyway"* (P8).

*Grounding risk identification with envisioned AI requirements.* Practitioners identified risks arising from the data AI products need for training or inference. P18 observed that AI Meeting Assistant

*"might expose more privacy-sensitive data like location, break time, cognitive load to accurately get the best time that the software can interrupt [a user]."* Others noted continuous data access (P21) or complex data flows: *"if it's using a third party service, then obviously that's like a whole another problem"* (P3).

*Grounding risk identification with prior privacy knowledge and experience.* Our participants also applied their own experience of privacy and grounded it with the **AI privacy taxonomy**. For example, P24 relied on their **experience as a user** to identify privacy risks through *"pattern matching... from prior social media experiences and what has been an issue with other platforms."* P14 relied on their **experience as a developer**: *"when I'm building AI products, this [distortion risk] is the highest quality issue we need to address. And you need to do a lot of evaluations to make sure this is at a safe level."*

*6.1.2 Privy helps practitioners brainstorm risk mitigation plans that are effective and appropriate.* On balance, our findings suggest that Privy was effective in helping participants to brainstorm mitigation strategies that are effective and appropriate in addressing the privacy risks they identified (DR3, DR4).

Participants reported high confidence in their mitigation plans. Of the 72 risks identified, participants rated their strategies as effective for 56 (78%), with most agreeing or strongly agreeing that their plans addressed the risks.

When evaluated by privacy experts, however, the mitigation plans received a more nuanced assessment. On a 6-point Likert scale (6 = strongly agree), experts rated the effectiveness of the plans *addressing identified risks* at an average of 3.86 (SD = 1.47), and 4.06 (SD = 1.47) for their *usefulness as conversation starters* in privacy discussions. For appropriateness, experts rated the plans a 4.21 (SD = 1.53) for *specificity to product concepts* and 3.92 (SD = 1.30) for *practicality*. Note that a 4 on this evaluation scale was labeled as "Slightly Agree" while a 3 was labeled as "Slightly Disagree".

Taken together, these results show that Privy enabled non-privacy-expert practitioners to articulate risk mitigation plans that could initiate broader privacy conversations, though additional support is needed to generate mitigation strategies that are immediately actionable and effective — a challenge we return to in Section 6.3.

This analysis also points to a notable gap in expert-assessed quality between the plans produced with $\text{Privy}_{LLM}$ versus $\text{Privy}_{Template}$ (Table 5). We explore the effects of the use of LLMs on the quality of privacy impact assessments in Section 6.2.

Our qualitative findings provided additional insight into how Privy helps participants create risk mitigation plans. Guided by Privy's *Privacy Risk Mitigator* (Section 4.3.1), when brainstorming risk mitigation plans, participants considered: (i) **the perspectives of the end user to address privacy barriers**, (ii) the **product utility-privacy tradeoff**, (iii) **end-user data management controls**, and (iv) **industry best practices**.

*Grounding risk mitigation in end-user perspectives.* Guided by the Privacy Risk Mitigator, participants grounded mitigation plans in addressing end-users' **awareness, motivation, and ability barriers** to the identified risks.

**To enhance end-users' awareness of privacy risks**, participants emphasized making data practices visible and transparent to end-users. For example, P3 suggested an *"AI meeting assistant should be implemented in a way that makes it visible, obvious to all attendees"* that their data is being collected and used.

**To enhance end-users' motivation to address privacy risks**, participants proposed strategies that would actively encourage end-users to reflect on and engage with their privacy-related choices. For example, P8 recommended showing users sample contacts excluded or included in Dynamic Audience Selection to *"encourage users to think twice, review their choices."*

**To enhance end-users' ability to manage privacy risks**, participants proposed strategies that would empower users with greater control and customization over how the AI in a product concept might be used. For example, for Dynamic Audience Selection, P20 proposed to *"give the option to the user to choose the topics that can be used to target them... [and] give an option to choose the people with whom they want to share their [AI-inferred] information."*

*Grounding risk mitigation in the utility-privacy tradeoff.* Some participants considered the utility-privacy tradeoff when brainstorming risk mitigation plans.

For instance, P7 articulated the **two sides of AI capabilities and privacy risks**, noting that eliminating certain risks may also remove the very utility that makes the AI valuable: *"you need audience graphs, social interaction, everything if you want to make the groups dynamically, if you want AI to generate the groups. But if you're defining groups [to reduce the risk], I don't think you need it [the product]."* Other participants described how **establishing guardrails was necessary** to prevent misuse while preserving core utility. For example, in the context of Dynamic Audience Selection, both P8 and P7 proposed restrictions that would prevent end-users from targeting sensitive or overly broad demographic groups.

*Grounding risk mitigation in end-user data management and control.* Some practitioners proposed mitigation strategies that centered on general best practices for secure handling of end-user data. These approaches were not unique to developing AI systems, but were envisioned to be essential nonetheless.

For example, P15 highlighted the need for a **secure data infrastructure**, suggesting that organizations invest in expertise to ensure privacy protections at the service level: *"why not just hire a person who has expertise on this... and make sure that this won't happen in the end?"* Others stressed **access control**, such as P4's proposal of *"strong conditional access policies... for employee access to data."* Participants also advocated for **data minimization**, as P21 noted: *"not everything is necessary... you don't necessarily need camera access and audio access, or can be enabled only at a specific time."*

*Grounded with industry best practices.* With Privy's scaffolding, some participants linked their mitigation strategies to industry standards that they know of. For example, P5 drew on design patterns from comparable products, noting that *"what Meta is doing... they've been through a similar situation about privacy. So their privacy center is really well designed."* Others emphasized regulatory compliance, such as P21's reference to Data Subject Access Request (DSAR):

**Table 5: Privacy expert evaluation on the effectiveness (addressing identified risks, useful conversation starter) and appropriateness (product specificity, practicality) of the proposed privacy risk mitigation plans.**

| Measurement: 6-point Likert Scale* | *Privy (Overall)* | *Privy$_{LLM}$* | *Privy$_{Template}$* |
|---|---|---|---|
| **Addressing Identified Risks** | 3.86 (SD=1.47) | 4.37 (SD=1.25) | 3.35 (SD=1.49) |
| **Useful Conversation Starter** | 4.06 (SD=1.47) | 4.69 (SD=1.16) | 3.42 (SD=1.50) |
| **Product Specificity** | 4.21 (SD=1.53) | 4.96 (SD=0.82) | 3.46 (SD=1.70) |
| **Practicality** | 3.92 (SD=1.30) | 4.42 (SD=1.10) | 3.42 (SD=1.30) |

*Note.* 1: strongly disagree, 2: disagree, 3: slightly disagree, 4: slightly agree, 5: agree, 6: strongly agree

*"which allows users to request their data.... companies should make sure that even their secondary data usage... is visible through forms."*

## 6.2 RQ2: How does the use of LLMs affect (i) the quality of privacy impact assessments produced by Privy, and (ii) the perceived usefulness of Privy in privacy risk envisioning and mitigation?

We next examined differences between Privy$_{LLM}$ and Privy$_{Template}$. First, we compared how privacy experts rated the quality of the privacy impact assessments produced by both Privy$_{LLM}$ versus Privy$_{Template}$ (Section 6.2.1). Then, we explored how practitioners perceived the usefulness of Privy$_{LLM}$ versus Privy$_{Template}$ in helping them complete their privacy impact assessments (Section 6.2.2). In both cases, we complemented our quantitative analysis with qualitative insights into participants' usage patterns and opinions towards Privy's LLM-powered features (Section 6.2.3).

*6.2.1 The use of LLMs improves the quality of privacy impact assessments produced by practitioners.* We evaluated the quality of privacy impact assessments produced with Privy$_{LLM}$ and Privy$_{Template}$ by analyzing the privacy risks identified and mitigation plans proposed by 24 participants (Privy$_{LLM}$: $N = 12$; Privy$_{Template}$: $N = 12$). Thirteen privacy experts independently rated these outputs on multiple quality measures (Fig. 6). Our ordinal logistic regression analyses revealed that participants using Privy$_{LLM}$ produced risk statements rated significantly higher in relevancy ($\beta = 1.95, p < .01$), severity ($\beta = 2.13, p < .001$), correctness ($\beta = 2.10, p < .001$), and clarity ($\beta = 2.39, p < .001$), compared to those generated with Privy$_{Template}$. A similar pattern emerged for risk mitigation plans: experts judged the Privy$_{LLM}$ group's mitigation plans to be significantly more effective — "addressing identified risks" ($\beta = 1.76, p < .001$) and "useful conversation starter" ($\beta = 2.51, p < .001$) — and appropriate — "product specificity" ($\beta = 2.04, p < .001$) and "practicality" ($\beta = 1.65, p < .01$). Overall, Privy$_{LLM}$ significantly outperformed Privy$_{Template}$ across all quality measures, including overall risk identification ($\beta = 2.45, p < .001$) and mitigation ($\beta = 1.95, p < .001$) quality, as well as the quality of the privacy impact assessment as a whole ($\beta = 2.45, p < .001$) (Fig. 7).

*6.2.2 The use of LLMs has minimal, if any, impact on the perceived usefulness and usability of Privy.* Beyond assessing output quality, we compared Privy$_{LLM}$ against Privy$_{Template}$ on practitioners' perceived usefulness and overall usability ratings across our five design goals (DG1-DG5). Both versions scored highly across these measures, indicating that each effectively helped users in achieving the intended goals (Fig. 8). System Usability Scale (SUS) scores were similarly strong for Privy$_{LLM}$ (M=76.04, SD=15.65) and Privy$_{Template}$ (M=75.63, SD=18.19), reflecting a "good" overall level of usability [9].

An ordinal logistic regression showed no significant differences between versions on perceived usefulness (**DG1:** $\beta = -0.98, p = 0.222$; **DG2:** $\beta = -0.15, p = 0.845$; **DG3:** $\beta = -1.38, p = 0.101$; **DG4:** $\beta = -0.44, p = 0.566$; **DG5:** $\beta = -0.12, p = 0.873$), and a linear regression on SUS scores likewise showed no significant difference ($\beta = 0.42, p = 0.953$). Taken together with Section 6.2.1, these findings suggest that while Privy$_{LLM}$ supports the production of higher-quality privacy impact assessments, practitioners perceived the usability and usefulness of both versions positively — we found no significant differences between the two versions.

*6.2.3 Use patterns and concerns with human-AI collaboration in privacy risk envisioning and mitigation.* We next explored how participants used the LLM-powered features in Privy$_{LLM}$, and their concerns therein.

*Exploring the unknown unknowns.* Participants frequently leveraged LLM-generated risk descriptions to **surface risks they had overlooked** or with which they were less familiar. As P7 describes: *"stuff like surveillance, risk intrusion, risk insecurity, I know when I see them... but it wouldn't come up [on my own], but I know when I see them, what it is, and it's actually risky."*

Participants also used LLM-generated examples to **compare and contrast risks**. For instance, P8 relied on LLM-generated risk descriptions to interrogate distinctions between disclosure and exposure risks, ultimately deciding that disclosure was more relevant to their AI product concept.

LLM-generated content helped participants articulate risk mitigation plans when they were unsure. P12 noted that the LLM's provocations were *"helpful when you don't have a line of thought to grasp on to"* during mitigation brainstorming.

*Solidifying and validating existing privacy knowledge.* Beyond surfacing new ideas, participants also turned to LLM-generated risks and provocations to **reinforce and cross-check** their apriori reasoning for identifying and mitigating privacy risks.

Some LLM-generated examples **helped improve and clarify practitioners' privacy knowledge**: *"because I'm not a privacy data practice practitioner... it helps me by outlining in more simple terms*

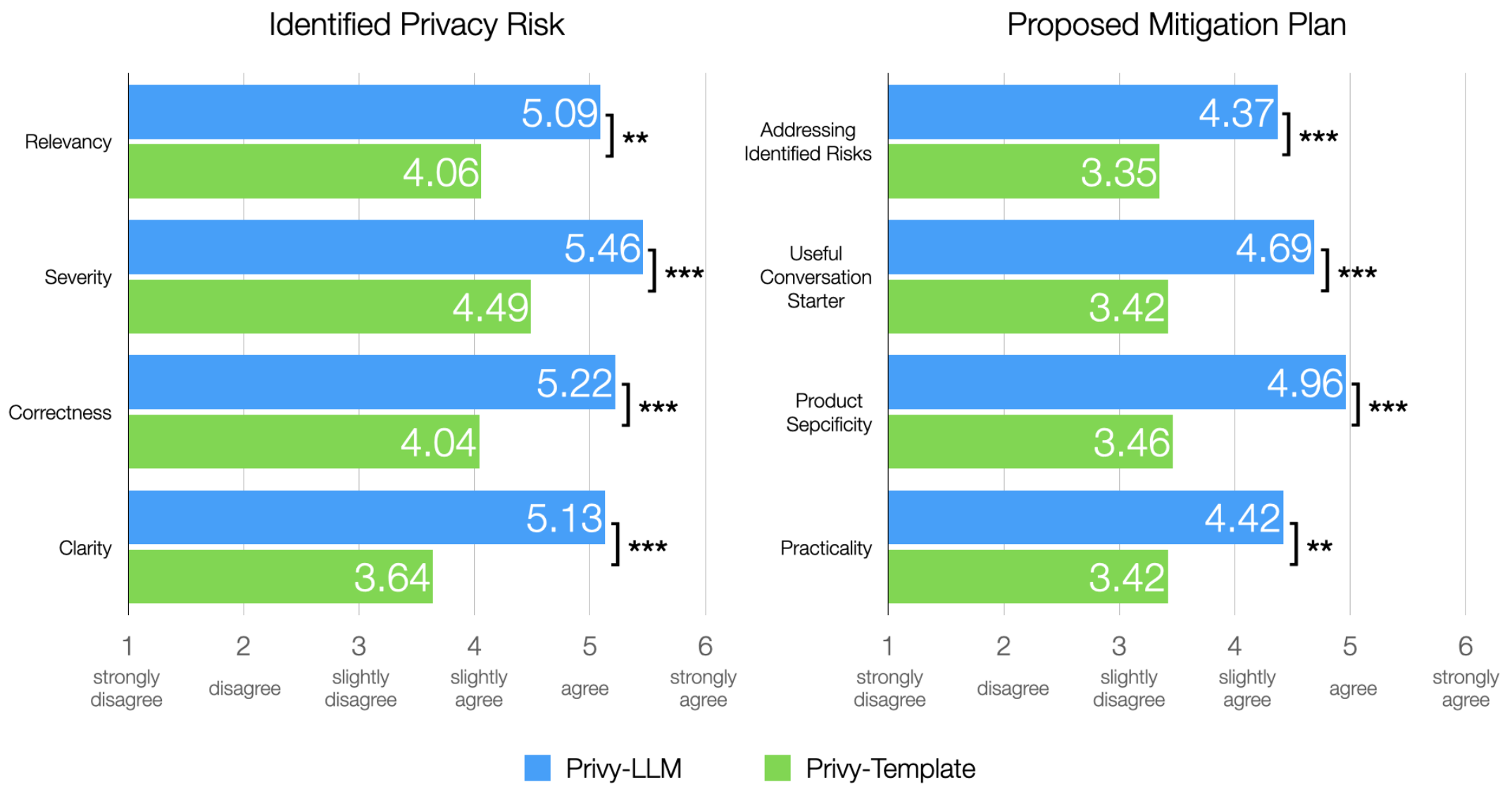

**Figure 6: Privacy expert ratings on the quality measurements of risk identification (left) and mitigation (right). Significance: *p<.05; **p<.01; ***p<.001**

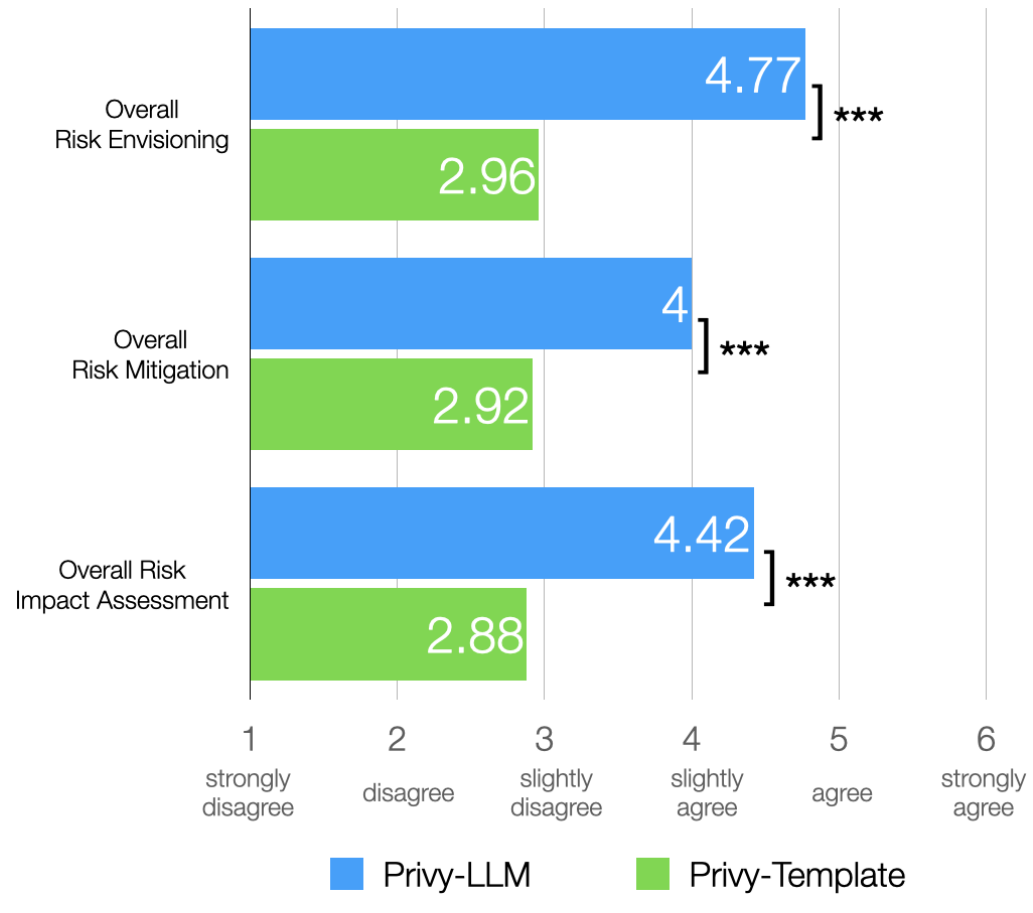

**Figure 7: Privacy expert ratings on the quality of the overall privacy impact assessment. Significance: *p<.05; **p<.01; ***p<.001**

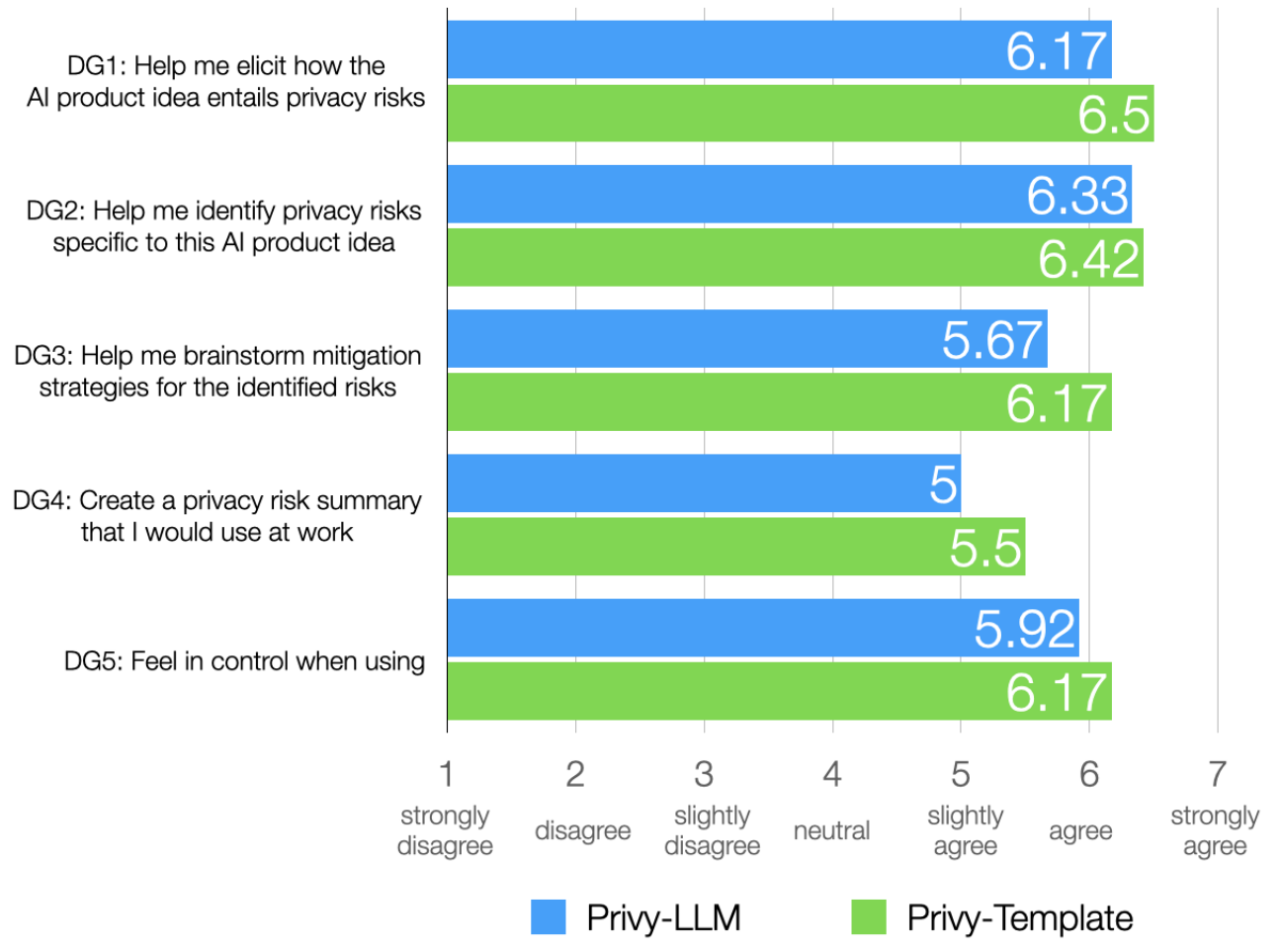

**Figure 8: Participants perceived that both Privy$_{LLM}$ and Privy$_{Template}$ were useful in achieving the design goals (DG1-DG5).**

*what the risk is and then how it impacts users"* (P11). Others critiqued LLM outputs against their own expertise. P1, for instance, dismissed generic suggestions like *"obtain consent"* as overly simplistic: *"you wouldn't set out to design a product without consent... and then pat yourself on the back when you add it."*

Some participants used Privy$_{LLM}$ as a thought partner, offering validation and reassurance. For example, as P3 put it, it served as **a sanity check on their reasoning** when working alone. P9 even described an "aha moment" when an LLM suggestion mirrored their own thinking: *"I had come up with something without having fully read the AI suggestion that the AI thought like, this is actually something you may want to consider."*

*Integrating LLM-generated ideas.* Participants often **integrated LLM-generated outputs** directly into their privacy impact assessments. In some cases, this meant accepting outputs as-is; in others, participants refined LLM suggestions. For example, P3 elaborated on a generated risk by detailing how it might happen: *"this could be a risk if it did the screenshot thing and then it was not on a protected server."* LLM outputs also served as **springboards for new,** **participant-driven ideas**. After reviewing LLM-generated use-cases, P8 identified an additional unintended use case for Dynamic Audience Selection: *"a user adds a minus marker, thinking it won't share with this group, but doesn't consider other parties and share it with them."*

LLM provocations helped participants **concretize abstract ideas** when articulating risk mitigation plans. As P2 shared: *"This is where we are heading to limited access itself. So the second one [provocation] would be helpful. I could add more stuff... mitigation steps over here based on the second question."*

*Concerns: tensions between automation and agency in privacy risk envisioning and mitigation.* Despite the benefits of LLM support, participants raised two main concerns: **overestimating the level of automated assistance provided by Privy$_{LLM}$**, and **the risk of overreliance on its outputs**. These concerns point to a core tension: while LLMs enhanced privacy assessment quality (Section 6.2.1), they also challenged practitioners' sense of agency in privacy work.

Some participants felt that parts of the workflow in Privy$_{LLM}$ (e.g., risk mitigation) were still driven primarily by their own effort and could be improved with more LLM-assistance. As P7 noted, *"I think the brainstorming mitigation tasks, I came up with them... I don't think I took so much help from Privy."* Others described the extra cognitive effort required to integrate AI suggestions into their privacy workflow: *"there's 7–12 different things [Privy$_{LLM}$-suggested risks]... that's a lot to juggle in my mind"* (P9).

Participants also worried that LLM-generated content could **bias their thinking** and **diminish cognitive engagement**. For example, P7 made a point to generate their own ideas before consulting Privy$_{LLM}$'s provocations: *"just making sure that doesn't create bias in my head."* Similarly, P1 emphasized pausing to reflect: *"have I included the risks that I felt were the most important ones before I started looking at the suggestions?"* P9 further warned that less experienced practitioners may be more prone to overreliance and uncritical acceptance: *"you may actually just be like, sure, this looks good, and accept."*

## 6.3 RQ3: What challenges do practitioners face in the privacy risk envisioning and mitigation process, and to what extent does Privy address these challenges?

In post-activity interviews, participants reflected on key challenges in envisioning and mitigating privacy risks for AI product concepts. We first synthesize these challenges (Section 6.3.1), drawing on the SPAF framework [19] to categorize them into barriers of **awareness**, **motivation**, and **ability**. We then examine how Privy — across both Privy$_{LLM}$ and Privy$_{Template}$ — helped participants address some of these barriers (Section 6.3.2), and where limitations remain (Section 6.3.3).

### 6.3.1 *Awareness, motivation, and ability barriers practitioners face in privacy risk envisioning and mitigation for AI product concepts.*

Guided by the SPAF, we uncovered awareness, motivation, and ability barriers practitioners faced when identifying and mitigating AI privacy risks.

*Awareness barriers.* Practitioners described the absence of a systematic **risk scanning mechanism**, particularly when products rely on multiple data flows and third-party integrations. As P20 explained, a key challenge was *"figuring out all the types of data you are taking from the user... but then you're using some third-party solution in your product, and they're using it, and you're not aware of it."*

Awareness barriers also stemmed from the challenge of anticipating real-world user behaviors. Participants noted the difficulty of **envisioning privacy implications from diverse user perspectives**, rather than relying solely on their own experiences. As P1 put it, the *"biggest challenge in brainstorming is having enough awareness of all the things that users do... you get very tunnel vision, focused into 'this is how I use products'... but you have to really think inside other users' minds and workflows."*

*Motivation barriers.* Echoing prior work [38], participants stressed that **privacy is rarely treated as a core priority**. P22 anticipated organizational resistance even if Privy was adopted: *"I can already imagine my boss will say... this is not our first priority. We don't need to worry about that... only if someone brings it up, or if the organization has a concern."* Some participants also described a tendency to **externalize responsibility**, framing privacy as the domain of specialized teams rather than part of their own remit. This distancing reduced ownership: unless privacy experts explicitly raised concerns, practitioners felt little obligation. As P6 noted, *"when these external teams don't tell us [about privacy issues], then we're not responsible for it."*

*Ability barriers.* Participants also struggled with their **ability to address privacy risks**, often framed as negotiating trade-offs between utility and intrusiveness. At the macro level, this required alignment across stakeholders. As P3 reflected: *"LLMs are obviously super useful... how do you want to make that trade-off? What kind of risk are you comfortable with? What risk is the executive leadership of your company acceptable with? It's kind of hard to quantify that sometimes."* At the micro level, trade-offs hinged on user experience and preferences, which were difficult to anticipate or accommodate: *"individual preference... the most challenging part, like how we accommodate variability in users' willingness to provide the data"* (P18).

Beyond trade-offs, some participants noted that they had no foundation to begin assessments. P8 described the hurdle of starting from scratch: *"the biggest hurdle would be when you start off on a blank page and you're like, 'Oh my God, where do I start?'"*

### 6.3.2 *Privy$_{LLM}$ and Privy$_{Template}$ address major barriers.*

We found that both Privy$_{LLM}$ and Privy$_{Template}$ could help practitioners address these longstanding awareness, motivation, and ability barriers in privacy work.

*Addressing awareness barriers: scaffolding structured privacy risk identification.* Privy's structured workflow provided critical scaffolding for envisioning privacy risks. Participants appreciated how the tool's concrete definition of privacy risks and taxonomy anchored their reasoning: *"understand the taxonomy here and then apply this taxonomy"* (P18). Others highlighted how the workflow supported prioritization, helping them discern underlying issues. As P17 reflected, *"comparing [risks] was fun... I started to realize* *maybe surveillance is not the big issue. Insecurity is the underlying cause."*

Privy's structured workflow enables practitioners to ground risk envisioning from various aspects (Section 6.1.1) and to reduce blind spots (Section 6.2.3), even without LLM-powered features. As P21 noted after using Privy$_{Template}$: *"there was a lot that came up that I had not considered previously."* Overall, Privy broadened many practitioners' awareness of privacy risks: 10 of 24 reported either perceiving more risks or perceiving fewer benefits of their AI product concepts after using the tool.

*Addressing motivation barriers: fostering reflection and engagement.* Privy encouraged participants to treat privacy as a reflective process. For instance, P9 noted that going through the *AI privacy risks assessment node* prompted them to refine their risk descriptions to better fit the product concept. Participants also emphasized that Privy made privacy assessments more enjoyable and generative, prompting deeper attention than they would normally give. As P20 explained: *"it's like the hacker mentality, right? You're trying to break things... in a privacy context. You can just brainstorm and come up with a lot of these examples. If you're actually doing it, you can spend like hours on this. I think it's enjoyable to me personally."*

*Addressing ability barriers: kickstarting self-efficacious privacy practice.* Echoing our quantitative results (Section 6.1), Privy enhanced practitioners' sense of self-efficacy in conducting privacy work. Participants reported feeling more capable and empowered after using the tool. As P24 reflected, Privy made them feel *"more confident that we have a solution here and we can move forward with a project idea,"* noting that surfacing both risks and corresponding solutions in one workflow *"gives you more control over the problem."*

This empowerment was reinforced by Privy's guided workflow, which nudged participants to broaden their consideration of risks. P4 described it as *"a little push... to say, hey, consider this thing as well."* Others valued the tool as a constructive entry point into collaborative privacy work: P8 envisioned using it in *"kickoff meetings or an early concepting phase of different ways we can use a particular AI tool."* These qualities — guidance and formative scaffolding — were among the features practitioners rated highly about Privy (Section 6.2.2).

*6.3.3 Limitations of Privy$_{LLM}$ and Privy$_{Template}$.* While both Privy$_{LLM}$ and Privy$_{Template}$ effectively supported practitioners in risk envisioning and mitigation, they were less effective for the latter (Section 6.1.2). Participants' feedback surfaced key design limitations, as well as pointed to opportunities for LLMs to more meaningfully support privacy impact assessments.

*More information needed for risk mitigation.* Privy offered a useful starting point, but participants stressed that effective mitigation planning requires a richer bank of information than the tool currently provides. P16 highlighted the need for concrete references: *"the example resolution that happened in both the backend side and the user side that I can refer to, then it can broaden my idea during the brainstorming."* Similarly, P18 pointed to industry best practices and examples of how similar products had addressed comparable risks: *"some examples [of] similar services like already launched. If that's the cases [potential risks], then I can learn from how they address that."* Beyond tool features, participants emphasized the broader need for education and training in privacy risk mitigation for AI development. As P6 put it, even with tools like Privy, practitioners remain responsible for making decisions and taking action: *"there needs to be some privacy training on using AI, which doesn't seem to exist very much right now."*

*Power dynamics in privacy decision-making.* Participants also described feeling constrained by organizational hierarchies around privacy decision-making. Even with tools like Privy, they felt their influence was limited in shaping outcomes. As P21 explained: *"I can bring up that there are privacy risks, but it's ultimately up to leadership to decide whether they abide by that, engineer around it, or still push the feature out."* These dynamics influenced how participants envisioned Privy's role. P7 felt the tool might be more relevant for compliance or security teams than developers: *"if developers use this, none of these reports would reach leadership, because I [as a developer] wouldn't want to get in trouble saying, hey, I missed this."* In response, some participants called for a more collaborative design that could engage multiple stakeholders. As P3 proposed: *"making it collaborative would be helpful... a lawyer would probably want to look at it, and my other teammates might also want to look at it."*

*LLMs that help people do privacy work.* Participants also envisioned alternative ways that LLM-powered features could better support privacy risk envisioning and mitigation. Some proposed dialogic interaction, where the system pushes them to refine and extend their thinking. P9 wanted Privy to *"help me refine some of the brainstorming, like having this back and forth of the dialog between me and the AI. In a developing team... it would be a dialog like people would be going back and forth."* Others, however, cautioned that LLMs should be designed to preserve and even strengthen practitioners' critical engagement with privacy work, a concern raised in Section 6.2.3. For example, P7 proposed delaying access to LLM-generated provocations until after users had articulated their own ideas, while P12 suggested limiting the number of AI-generated suggestions offered at once.

These perspectives highlight an opportunity for more nuanced, deliberately scaffolded forms of LLM support, helping practitioners generate ideas and improve their ability in privacy decision-making. We return to this opportunity in Section 7.2.

# 7 Discussion

## 7.1 Empowering AI practitioners to address privacy risks

Prior work has highlighted how the *lack of practical tools* is a key barrier preventing practitioners from systematically addressing privacy risks in AI products [38]. To address this gap, we developed Privy, grounded in the needs and insights of AI and privacy practitioners (Section 3), to support the identification and mitigation of privacy risks in AI product concepts. Our findings show that both Privy$_{LLM}$ and Privy$_{Template}$ enabled non-privacy-expert practitioners to produce privacy impact assessments rated highly by privacy experts (Section 6.1) and perceived as useful and usable by practitioners themselves (Section 6.2.2). More broadly, Privy empowers practitioners to envision and mitigate privacy risks — addressing challenges of awareness, motivation, and ability that often impede privacy work (Section 6.3.2).

From Privy's development and evaluation, we distilled three key shifts that future tools should aim to help non-privacy-expert practitioners meaningfully engage with privacy work in the age of AI.

***From a passive and generic privacy education, to proactive and product-driven privacy case studies.*** Our findings reaffirm that practitioners' privacy knowledge underpins their **awareness** and **ability** to address risks. Yet, echoing prior work [38], participants described industrial training as overly generic, rarely tailored to AI, and disconnected from their specific product contexts. Privy, by contrast, scaffolded more rigorous and product-relevant engagement with privacy (Section 6.3.2). Practitioners reported that using Privy heightened their awareness of privacy issues and inspired them to seek additional resources when developing mitigation plans (Section 6.3.3). Many grounded their responses in industry best practices and expressed a desire for more relevant examples (Sections 6.1.2 and 6.3.3).

These findings suggest that *product-driven privacy case studies* can complement traditional privacy training. Future tools should integrate relevant case studies, industry best practices, and real-world incidents directly into practitioners' workflows. Researchers can extend our design strategies by embedding such case studies within interfaces like Privy's AI Capability & Requirement Scaffolder, Risk Explorer, and Risk Mitigator to further support practitioners in envisioning and mitigating privacy risks in their design process.

***From supporting privacy novices to cultivating privacy champions.*** Practitioners often face organizational resistance when privacy work appears to be misaligned with product team goals — an issue surfaced in both our findings (Section 6.3.3) and prior work [38]. Our findings suggest that a promising path forward is to make privacy work inherently collaborative, by involving non-privacy-expert stakeholders in privacy discussions and decision-making early in product development (Sections 6.3.2 and 6.3.3). Beyond collaboration, our findings indicate that tools like Privy can help practitioners move past compliance-driven tasks toward becoming *privacy champions* within their teams [38, 63]. With Privy, participants described feeling ownership and motivation to invest more effort in privacy work, rather than treating it as just another worksheet (Section 6.3.2). Importantly, many produced privacy impact assessments of sufficient quality to engage privacy experts in conversations around product-specific privacy risks (Sections 6.1.1 and 6.1.2) — an early signal that tools like Privy can re-balance privacy decision-making power.

Future tools should continue to support these **motivational** shifts — helping practitioners advocate for privacy internally. Beyond generating assessments, systems could scaffold how findings are shared: presentations in team meetings or summaries for leadership. In doing so, Privy-like tools could foster a culture where practitioners not only participate in privacy work, but take a leading role.

***From fixed utility-privacy tradeoffs to nuanced utility-privacy configurations.*** Consumer-facing AI products are often built around a tradeoff: users gain full access to services in exchange for relinquishing private data [1, 41]. Yet our findings suggest that practitioners envision more nuanced approaches — enabling end-users to tailor AI functionality based on their privacy preferences, such as limiting specific capabilities. Participants also acknowledged the challenges of this approach, including ensuring product reliability across varying data availability (Sections 6.1.2 and 6.3.1). When supported by tools like Privy, practitioners were able to generate solutions that moved beyond company-centric (often incorrect) assumptions, such as "users do not care about their privacy settings" [43, 60], toward designs that foreground user agency. By scaffolding reflection on privacy from varied end-users' perspectives, Privy helped practitioners articulate options that empower users to take greater control of their personal data (Section 6.1.2).

Future work can build on this success by developing *user-defined, privacy-preserving AI products*, where individuals actively configure the utility–privacy balance of their experience. Privy can be extended to guide practitioners in designing such configuration mechanisms, ensuring that tradeoffs are not imposed unilaterally but negotiated through interfaces that respect user choice.

## 7.2 Generative AI-powered privacy risk envisioning and mitigation

In this study, we examined how generative AI (GenAI) can support practitioners in envisioning and mitigating privacy risks (RQ2). Our results are promising: augmenting Privy with an LLM ($\text{Privy}_{LLM}$) enhanced its effectiveness in helping practitioners identify and mitigate privacy risks (Section 6.2.1). Our findings also showed how integrating GenAI could generate higher-quality privacy impact assessments and enable practitioners to engage critically in the assessment process itself (Section 6.3.3).

Building on these insights, we articulate three design objectives for embedding GenAI-powered tools into non-privacy-expert practitioners' workflows to better support their privacy risk envisioning and mitigation practices.

***GenAI-powered tools as privacy coaches.*** Privacy education is increasingly critical in the AI era, where emerging capabilities often introduce novel and unforeseen privacy harms [40]. Our findings suggest that GenAI-powered tools like $\text{Privy}_{LLM}$ can act as *privacy coaches*, training practitioners as they work — helping them reason through risks and understand key concepts in context (Sections 6.2.3 and 6.3.2).

These tools can introduce foundational frameworks (e.g., the AI privacy taxonomy [40]) through product-specific examples, reinforcing concepts within the practitioner's workflow. Extending our earlier notion of *product-driven privacy case studies* (Section 7.1), GenAI can dynamically surface relevant examples, case studies, and resources as practitioners progress through assessments. An added benefit is adaptability: unlike fixed training modules, GenAI-powered tools can keep privacy guidance current — reflecting rapidly shifting technological and regulatory landscapes. This positions GenAI not just as a knowledge source, but as an ever-learning privacy coach embedded in practice.

***GenAI-powered tools as privacy co-pilots.*** One of GenAI's most immediate contributions to privacy work is its role as a *co-pilot* — a thought partner that enhances both the quality and efficiency of privacy impact assessment. Our findings show that GenAI helped participants think beyond their blind spots and sped up tasks like pre-populating risk descriptions — features they found especially valuable (Section 6.2).

As co-pilots, GenAI-powered tools can enable richer forms of human–AI collaboration in privacy risk envisioning and mitigation. Beyond facilitating early-stage discussions and decision-making, these tools could be extended to be more prescriptive in suggesting risks and potential solutions (e.g., engineering best practices), support hypothesis testing, or help teams interrogate design trade-offs. Future systems may compare risks across alternative technical implementations or simulate end-user behaviors — especially privacy impacts on marginalized groups — to prompt practitioners to refine product concepts. Such capabilities may enable forms of design iteration and exploration that are difficult to achieve manually.

Looking ahead, comprehensive evaluation of the *quality* of AI-assisted assessments remains essential as suggestions become increasingly product-specific and actionable. For example, it would be prudent to create a ground truth benchmark to measure alignment between LLM-generated and expert privacy impact assessments. At the same time, future research should go beyond evaluating assessment quality to examine how GenAI-augmented workflows shape actual product decision-making.

***GenAI-powered tools as privacy provocateurs.*** While participants valued GenAI-generated content, they also emphasized the need to maintain agency and stay critically engaged in their privacy work (Sections 6.2.3 and 6.3.3). This observation aligns with prior work showing that critical engagement is heightened when practitioners have confidence in their own work when using GenAI tools [39] — a finding echoed in our study as our participants expressed high self-efficacy when using Privy$_{LLM}$ (Section 6.1).

Positioned as *provocateurs*, GenAI-powered tools should foster long-term growth in privacy reasoning, not just supply answers [54]. These tools should be designed to nudge practitioners toward reflection, prompting them to refine, challenge, or reimagine their own ideas [15, 66]. In this role, GenAI acts not as a substitute for privacy expertise but as a catalyst for deeper engagement.

This provocateur role is especially vital in contexts requiring nuanced privacy decisions — such as the *utility–privacy configurations* discussed earlier (Section 7.1) — where there are no clear right answers. Here, GenAI-powered tools can help practitioners navigate competing design possibilities and trade-offs, ultimately guiding them toward confident, well-reasoned decisions.

## 7.3 Limitations

Our work has several limitations. In the evaluation study, participants assessed two researcher-provided AI product concepts rather than their own. This decision enabled a controlled, between-subject comparison of Privy$_{LLM}$ and Privy$_{Template}$ (RQ2) and ensured feasible completion within the 90-minute session. While we ensured participants understood the assigned concepts, their engagement with privacy risks may differ when working on their own products.

Time constraints also shaped our study design. We allocated 25 minutes for risk identification and 15 minutes for mitigation — enough to complete the activities, but shorter than real-world, iterative development cycles. As such, our findings reflect a snapshot of tool use under structured conditions. Future work could involve longitudinal deployments of tools like Privy with real product teams to observe sustained patterns of use and adoption. Such studies may shed light on how Privy shapes privacy-related product decision-making and business practices over time and how it integrates with — or reshapes — existing privacy impact assessment workflows.

Additionally, in real-world deployments, product teams using Privy-like systems could leverage more advanced LLMs and fine-tune underlying models to better align with their domain and organizational context — beyond the few-shot configuration used in our study. As such, the effectiveness of Privy can be further enhanced in practice.

Finally, our participant pool was limited to practitioners based in the US and the UK. Although our 35 participants from industry (across both the formative study and summative evaluation) spanned diverse roles and organizations, our findings may not fully generalize across all work contexts.

# 8 Conclusion

Privy is a novel privacy risk-envisioning tool that helps AI practitioners without privacy expertise identify and mitigate the most relevant and severe privacy risks posed by their AI product concepts. In a controlled experiment with 24 non-privacy-expert practitioners, whose Privy-scaffolded outputs were evaluated by 13 independent privacy experts, we showed that Privy is both useful and usable, and enables practitioners to generate high-quality privacy impact assessments. Privy ensures privacy impact assessments are both significant and contextual by grounding risk envisioning and mitigation in product-specific AI capabilities & requirements, plausible use-cases, and stakeholders who are likely to be impacted. Our findings further show that, with thoughtful friction, generative AI can enhance the risk envisioning and mitigation process: participants used LLM-generated content to surface overlooked risks, validate their reasoning, and expand mitigation strategies. This work opens new directions for designing tools that empower AI practitioners to meaningfully engage with privacy, and for critically shaping the roles generative AI can play in supporting privacy-preserving AI innovation.

# Acknowledgments

This work was generously supported, in part, by NSF SaTC grants #2316768 and CyLab Presidential Fellowship. We also thank our anonymous reviewers for being generous with their time and helping us improve the paper.

## A PRIVY$_{LLM}$ Privacy Risk Summarizer Interface

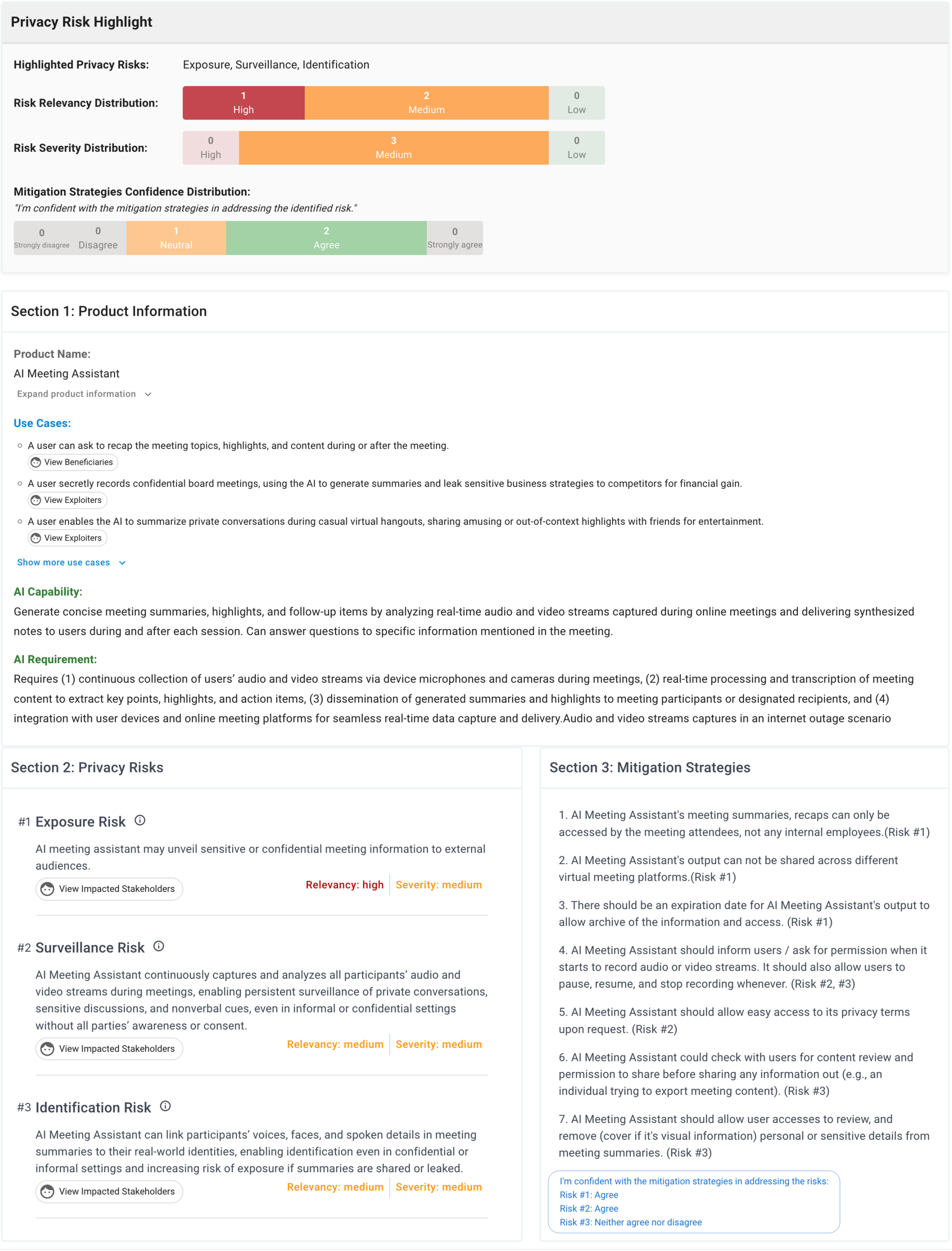

**Figure 9: Privacy Risk Summarizer: the privacy impact assessment report summarizes practitioners' product description, use cases, AI capabilities and requirements (Section 1: Product Information), identified privacy risks (Section 2: Privacy Risks), and proposed mitigation strategies (Section 3: Mitigation Strategies).**

# B PRIVY$_{Template}$ Interface

## AI Privacy Template

**Product Name:**

**Product Purpose:**

**Product Data:**

### Section 1: Product Information

**Potential Use Cases and Users:** *1) What potential intended or unintended use cases could arise with this product, and 2) who are the entities or individuals that will use and benefit from the product?*

- **Use Case 1 | Select Type :** A user uses the product to...
  - *User(s):*
- **Use Case 2 | Select Type :**
  - *User(s):*
- **Use Case 3 | Select Type :**
  - *User(s):*
- **Use Case 4 | Select Type :**
  - *User(s):*

**AI Capabilities:** *What can the AI do?*

-
-

**AI Requirements:** *What data or infrastructure is needed for this AI to work?*

-
-

### Section 2: Privacy Risks

*Step 1 Identify: Identify privacy risks entailed by the product (1.1), their associated impacted stakeholders (1.2), and their relevance and severity (1.3).*
*Step 2 Rank: Rank the order (1st - 3rd) in which you think these risks should be addressed.*

***Relevancy:*** *Considering the product purpose, this risk is relevant to the AI user experience:*

- *High: The risk is highly relevant and closely tied to the use cases and product.*
- *Medium: The risk is somewhat relevant to the use cases and product, though the relevance might be indirect and speculative in certain situations.*
- *Low: The risk has minimal direct relevance to the use cases and product. In rare instances, it may have indirect relevance (e.g., speculated downstream impact).*

***Severity:*** *Considering the product purpose, the risk poses a significant impact on society and/or individuals due to the AI:*

- *High: The risk is substantial and directly harmful on an individual and/or societal scale.*
- *Medium: The risk is noticeable and harmful at the individual level and/or societal level.*
- *Low: The risk is negligible at both the individual and societal scales.*

| **Privacy Risk #** | Step 1.1: Describe the privacy risks you think might arise from this product. *You can refer to the taxonomy and describe how the product can cause a privacy risk.* | Step 1.2: List the potential impacted stakeholders of the risk. | Step 1.3: Rate the relevancy and severity of the risk. | Step 2: In what order do you think these risks should be addressed? |
|---|---|---|---|---|
| **#1** | Risk type: *Please select*<br>Risk description: | | Relevancy score: *Please select*<br>Severity score: *Please select* | Please select |
| **#2** | Risk type: *Please select*<br>Risk description: | | Relevancy score: *Please select*<br>Severity score: *Please select* | Please select |
| **#3** | Risk type: *Please select*<br>Risk description: | | Relevancy score: *Please select*<br>Severity score: *Please select* | Please select |

### Section 3: Mitigation Strategies

*Outline mitigation strategies associated with the three privacy risks you identified (3.1), and rate how confident you are these strategies address those risks (3.2).*

Step 3.1:
Considering all three privacy risks, outline how you would mitigate the risks to create a privacy-preserving version of the product. Try to be as specific as possible.

*For each risk, you might think about:*

- *How can the product be designed to enhance users' awareness of the risk?*
- *How can the product be designed to enhance users' motivation to address the risk?*
- *How can the product be designed to provide users with the ability to manage the risk?*

Step 3.2:
I'm confident in the proposed mitigation strategies for addressing this risk.

*Please rate the degree to which you agree with this statement for each identified risk.*

| **Privacy Risk #** | |
|---|---|
| **#1** | *Please select* |
| **#2** | *Please select* |
| **#3** | *Please select* |

**Figure 10: The complete worksheet of PRIVY$_{Template}$.**

## Reference: AI Privacy Risk Taxonomy

| | | |
|---|---|---|
| **Data Collection Risks** *create disruption based on the process of data gathering* | | |
| | Surveillance (see examples) | Watching, listening to, or recording of an individual's activities. |
| **Data Processing Risks** *result from the use, storage, and manipulation of personal data* | | |
| | Identification (see examples) | Linking specific data points to an individual's identity. |
| | Aggregation (see examples) | Combining various pieces of data about a person to make inferences beyond what is explicitly captured in those data. |
| | Phrenology & Physiognomy (see examples) | Inferring personality, social, and emotional attributes about an individual from their physical attributes. |
| | Secondary Use (see examples) | The use of personal data collected for one purpose for a different purpose without end-user consent. |
| | Exclusion (see examples) | The failure to provide end-users with notice and control over how their data is being used. |
| | Insecurity (see examples) | Carelessness in protecting collected personal data from leaks and improper access due to faulty data storage and data practices. |
| **Data Dissemination Risks** *result when personal information is revealed or shared by data collectors to third parties* | | |
| | Exposure (see examples) | Revealing sensitive private information that people view as deeply primordial that we have been socialized into concealing. |
| | Distortion (see examples) | Disseminating false or misleading information about people. |
| | Disclosure (see examples) | Revealing and improperly sharing data of individuals. |
| | Increased Accessibility (see examples) | Making it easier for a wider audience of people to access potentially sensitive information. |
| **Invasion Risks** *the unwanted encroachment into an individual's personal space, choices, or activities* | | |
| | Intrusion (see examples) | Actions that disturb one's solitude in physical space. |

**Figure 11: Reference of Privy$_{Template}$: AI privacy taxonomy [40].**

# C Formative Study Appendix

## C.1 Privy$_{beta}$'s Interfaces

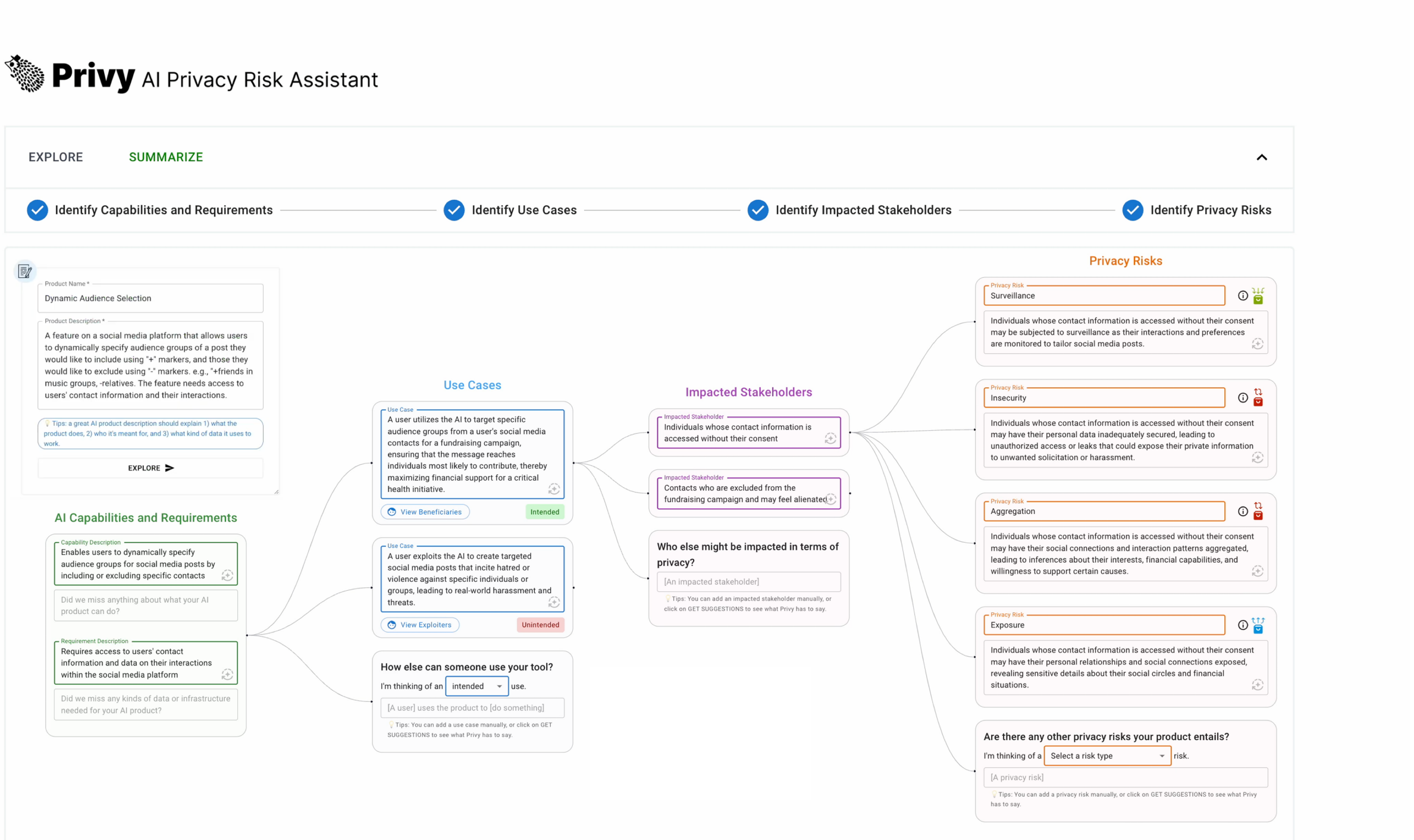

**Figure 12: Risk explorer: a multi-step scaffold that helps users articulate the purpose, use cases, and impacted stakeholders of an AI product concept, and guides them to envision potential privacy risks based on the AI privacy taxonomy.**

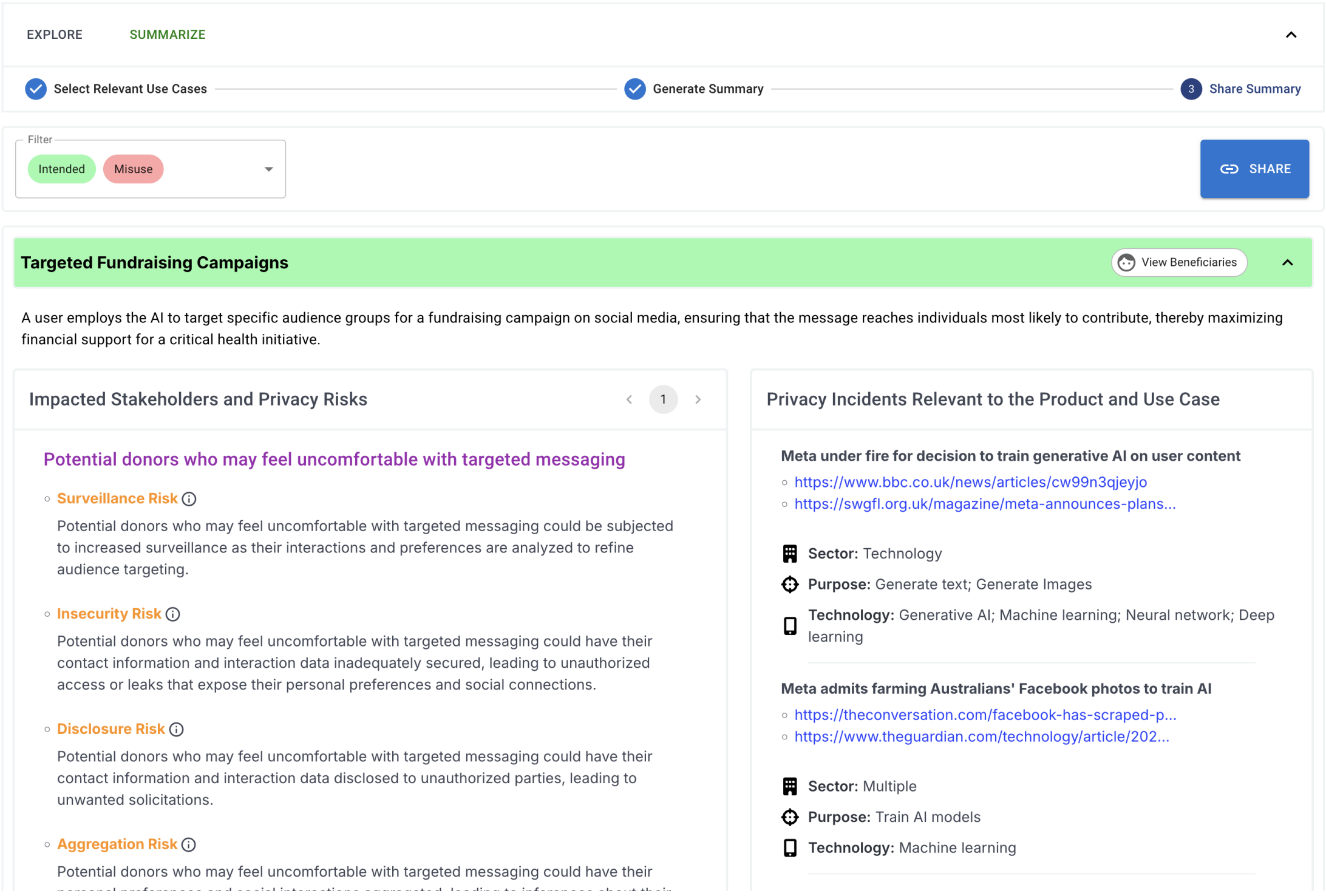

**Figure 13: Risk summarizer: synthesizes the privacy risks identified through the risk explorer, links them to relevant real-world privacy incidents, and generates a shareable summary.**

## C.2 Formative Study Participants

**Table 6: The formative study included 11 AI and privacy practitioners with diverse job roles.**

| Participant Roles | Participant IDs |
|---|---|
| Research Scientist | U7, U9 |
| Software Engineer | U1, U2, U3, U6 |
| Privacy Engineer | U10, U11 |
| Product Manager | U8 |
| Product Designer | U4, U5 |

## C.3 Formative Study Interview Questions

(1) Pre-tool use warm-up questions
  (a) What is the AI product or idea that you and your team are currently working on?
    (i) Who are the end-users of the product?
  (b) Can you think of the last time you and your team discussed privacy risks with respect to this product?
    (i) (If yes) During that discussion, how did you and your team identify the privacy risks of your AI product?
      (A) Can you share the types of privacy risks that you considered?
      (B) If, and how did you and your team act after identifying those privacy risks?
    (ii) (If no)
      (A) Who in your team takes care of privacy considerations?
      (B) What is your working relationship with [the person]?
(2) Tool feedback session: risk explorer
  (a) Please rate and elaborate on the value of each feature in identifying privacy risks of your AI product idea. Rate each feature using a 5-point Likert scale with the statement: "The feature brings value for my privacy work." - Strongly disagree (1), Disagree (2), Neither agree nor disagree (3), Agree (4), Strongly agree (5), elaborate on the reasons why:
    (i) capability/ requirement summary node
    (ii) use cases node
    (iii) beneficiaries/ exploiters
    (iv) impacted stakeholders node
    (v) privacy risk vignettes node
  (b) Compared with the privacy risks you and your team identified before using our tool, if and how does the tool prompt you to identify new risks that you've not thought of?
  (c) How would the tool's features prompt you to think or do differently in identifying potential privacy risks of your product idea?
    (i) Who in your team would be interested in the information provided by the tool?
(3) Tool feedback session: risk summarizer
  (a) Please rate and elaborate on the value of each feature in identifying privacy risks of your AI product idea. Rate each feature using a 5-point Likert scale with the statement: "The feature brings value for my privacy work." — Strongly disagree (1), Disagree (2), Neither agree nor disagree (3), Agree (4), Strongly agree (5), elaborate on the reasons why:
    (i) AI utility-privacy trade-off summary
    (ii) relevant real-world AI privacy incidents
    (iii) export report
  (b) How would the tool prompt you to think or do differently about informing discussions and decisions about privacy?
    (i) Who in your team would be interested in the information provided by our tool?
(4) Post-tool use interview
  (a) Can you think of a discussion your team had about how to design [participant's product] that may have gone differently if you had the tool's outputs?
    (i) Specifically, if and how these outputs are useful for you and your team in that discussion?
      (A) (If yes) Why would the outputs be useful?
      (B) (If yes) Who on your team would you want to share these results with?
      (C) (If no) Why would the outputs be unuseful?
      (D) (If no) How would the information be useful for you and your team?
    (ii) More broadly, how do you think the tool might fit into your and your team's typical AI product life cycle?
    (iii) Let's imagine together, if you have a magic wand to change anything:

(A) How would you change the tool's design to better support your ability to identify potential privacy risks?
(B) How would you change the tool's design to better inform your team's discussions and decisions about privacy?
(iv) Finally, before we close this interview, do you have any other thoughts you would like to share about our tool, or your experience in doing privacy work?

# D Evaluation Study Appendix

## D.1 Evaluation Study Participants

**Table 7: The evaluation user study included 24 participants with diverse job roles.**

| Participant Roles | Participant IDs |
|---|---|
| Research Scientist | P3, P6, P7, P9, P12, P18, P24 |
| Software Engineer | P1, P13, P14, P15, P19 |
| Machine Learning Engineer | P10, P21 |
| Security Engineer | P4, P20 |
| Product Designer | P5, P8, P11, P17, P22 |
| UX Researcher | P2, P16, P23 |

## D.2 Evaluation Study Interview and Survey Questions

### *D.2.1 Interview Questions.*

(1) Pre-activity warm-up questions
(a) What is the AI product or idea that you and your team are currently working on?
(b) Can you think of the last time you and your team discussed privacy risks with respect to this product?
(i) (If yes) During that discussion, how did you and your team identify the privacy risks of your AI product?
(A) How comfortable are you approaching potential privacy concerns in the products you're developing?
(ii) (If no) Why not?
(2) During-activity follow-up questions
(a) Practitioners' approaches to identifying privacy risks
(i) Could you walk me through how you identify the risk?
(ii) Could you walk me through how you assess the risk?
(b) Practitioners' approaches to brainstorming mitigation strategies
(i) Can you walk me through how you came up with mitigation strategies for the privacy risk?
(ii) Could you elaborate on your confidence in your mitigation strategies for the risk?
(iii) What are your thoughts on the brainstorming questions for this risk?
(c) Debriefing
(i) Does Privy change anything about how you approach or think about privacy?
(ii) Seeing this privacy summary report once again, will you be confident in presenting these findings to your team?
(iii) Is there anything else you would like us to know about your experience using the tool?
(3) Post-activity interview questions
(a) What is your thought on developing [the assigned product concept] now after engaging in this exercise?
(b) Were there any features or designs of Privy that stood out to you? Why?
(c) Were there any challenges you encountered while using the tool? If so, can you describe them?
(d) If you had a magic wand and could change anything about Privy, what would it be?
(e) Now, let's step back and discuss your overall challenges in privacy work
(i) What is your biggest challenge in identifying product-specific privacy risks?
(ii) What is your biggest challenge in brainstorming mitigation strategies for privacy risks?

### *D.2.2 Post-activity survey questions.*

(1) Please rate the degree to which you agree or disagree with said statements. (7-point Likert scale: strongly disagree, disagree, slightly disagree, neither agree nor disagree, slightly agree, agree, and strongly agree)
(a) Privy helped me elicit how the AI product idea entails privacy risks.
(b) Privy helped me identify privacy risks specific to this AI product idea.
(c) Privy helped me brainstorm mitigation strategies for the identified risks.

(d) Privy created a privacy risk summary that I would use at work.
   (e) I felt in control when using Privy.
(2) *(System Usability Scale)* Please rate the degree to which you agree with the following statements. (5-point Likert scale: strongly disagree, disagree, neither agree nor disagree, agree, and strongly agree)
   (a) I think that I would like to use this system frequently.
   (b) I found the system unnecessarily complex.
   (c) I thought the system was easy to use.
   (d) I think that I would need the support of a technical person to be able to use this system.
   (e) I found the various functions in this system were well integrated.
   (f) I thought there was too much inconsistency in this system.
   (g) I would imagine that most people would learn to use this system very quickly.
   (h) I found the system very cumbersome to use.
   (i) I felt very confident using the system.
   (j) I needed to learn a lot of things before I could get going with this system.

## D.3 Codebook for the Evaluation Study Qualitative Analysis

### *D.3.1 RQ1: How and to what extent do Privy$_{LLM}$ and Privy$_{Template}$ help practitioners identify and mitigate privacy risks with AI product concepts?*

(1) **How Privy affects practitioners' risk identification approach**
   (a) Grounding risk identification with use cases
      - *Use case-driven privacy risk manifestation*
      - *Privacy risks of indirect stakeholders*
   (b) Grounding risk identification with envisioned AI requirements
      - *Threat modeling: data sensitivity*
      - *Threat modeling: the ubiquity and the scale of data collection*
      - *Threat modeling: data flow and leakage*
   (c) Grounding risk identification with envisioned AI capabilities
      - *Privacy implications embedded with the envisioned AI capability*
      - *Downstream AI usage risk*
      - *Failure mode-entailed privacy risk (the AI does not perform as expected)*
   (d) Grounding risk with AI privacy taxonomy
   (e) Grounding risk identification with prior privacy knowledge and experience
      - *Personal experiences as a user*
      - *Attitudes toward certain risks*
(2) **How Privy affects practitioners' risk mitigation approach**
   (a) Grounding risk mitigation in end-user perspectives
      - *Enhance end-users' awareness of privacy risks*
      - *Enhance end-users' motivation to address privacy risks*
      - *Enhance end-users' ability to manage privacy risks*
   (b) Grounding risk mitigation in the utility-privacy tradeoff
      - *Consider the two-side of AI capabilities and privacy risks*
      - *Establish guardrails to prevent misuses*
   (c) Grounding risk mitigation in end-user data management and control
      - *Secured data infrastructure*
      - *User data protection and access control*
      - *Data minimization*
   (d) Grounded with industry best practices

### *D.3.2 RQ2: How does the use of LLMs affect (i) the quality of privacy impact assessments produced by Privy, and (ii) the perceived usefulness of Privy in privacy risk envisioning and mitigation?*

(1) **Use patterns with human-AI collaboration in privacy risk identification and mitigation**
   (a) Come up with a diversified set of risks
   (b) Catch overlooked risks
      - *Explore privacy implications of use cases, stakeholders, and product details*
      - *Explore the unknown unknown*
   (c) Compare and contrast different risks
   (d) Solidifying and validating existing privacy knowledge

- *Criticize the LLM-generated examples against their own privacy knowledge*
- *Use LLM-generated examples to improve privacy knowledge*

(e) Integrated LLM-generated outputs
- *Add details of potential risk manifestation*
- *Add relevant stakeholders*
- *Concretize abstract ideas*

(2) **Users' concerns with human-AI collaboration in privacy risk identification and mitigation**

(a) Perceived a need to spend more efforts than expected

(b) Concerns about overreliance on Privy-LLM
- *Thoughts being biased*
- *Diminish cognitive engagement*

#### D.3.3 RQ3: What challenges do practitioners face in the privacy risk envisioning and mitigation process, and to what extent does Privy address these challenges?

(1) **Challenges in practitioners' privacy work**

(a) Awareness barriers
- *Lack of a systematic risk scanning mechanism*
- *Difficulty in envisioning privacy implications from diverse user perspectives*
- *The need to reference industry best practices and examples*

(b) Motivation barriers
- *Privacy is not a priority*
- *Perceived the need to have privacy handled by others*

(c) Ability barriers
- *Balance the utility and intrusiveness for AI products and services*
- *Not knowing how to start tackling privacy risks*

(2) **Feedback on Privy**

(a) Addressing awareness barriers
- *Provide a clear categorization/definition of privacy risks*
- *Enhance the awareness of the importance of some risks*
- *Enhance overall privacy risk awareness*

(b) Addressing motivation barriers
- *Foster reflection and engagement in privacy work*

(c) Addressing ability barriers
- *Enhance perceived self-efficacy of privacy work*
- *Provide a guided workflow for privacy work*
- *Provide a good starting point for collaborative privacy work*

(3) **Privy's Limitations**

(a) How Privy can be improved
- *More information needed for risk mitigation*
- *Power dynamics in privacy decision-making*
- *Make the process collaborative*

(b) How the integration of LLM can be improved
- *Help improve practitioners' risk envisioning and mitigation*
- *Introduce mechanisms to help deter overreliance on AI*

## D.4 Privacy Risk Types Identified by Practitioners and Privacy Experts

**Table 8: We found that both PRIVY$_{LLM}$ and PRIVY$_{Template}$ enabled practitioners to identify a broad spectrum of privacy risks across both AI product concepts. Overall, the risk types they surfaced aligned with those identified by the privacy experts. While practitioners identified a slightly more diverse set of risk types using PRIVY, this variation should be interpreted tentatively given the study's limited sample size.**

| | AI Meeting Assistant | | | | Dynamic Audience Selection | | | |
|---|---|---|---|---|---|---|---|---|
| *Counts of Each Risk Type Selected in PIA Reports* | PRIVY$_{Template}$ | PRIVY$_{LLM}$ | PRIVY (Overall) | Privacy Experts* | PRIVY$_{Template}$ | PRIVY$_{LLM}$ | PRIVY (Overall) | Privacy Experts* |
| Surveillance | 6 | 3 | 9 | 7 | 2 | 6 | 8 | 2 |
| Identification | 3 | 1 | 4 | 0 | 3 | 2 | 5 | 3 |
| Aggregation | 2 | 3 | 5 | 1 | 2 | 0 | 2 | 4 |
| Phrenology Physiognomy | 0 | 0 | 0 | 0 | 0 | 0 | 0 | 0 |
| Secondary Use | 2 | 1 | 3 | 3 | 1 | 0 | 1 | 2 |
| Exclusion | 0 | 2 | 2 | 4 | 2 | 0 | 2 | 0 |
| Insecurity | 2 | 1 | 3 | 1 | 1 | 1 | 2 | 1 |
| Exposure | 1 | 3 | 4 | 0 | 1 | 2 | 3 | 1 |
| Distortion | 1 | 2 | 3 | 3 | 1 | 2 | 3 | 2 |
| Disclosure | 1 | 0 | 1 | 1 | 2 | 2 | 4 | 3 |
| Increased Accessibility | 0 | 1 | 1 | 1 | 2 | 0 | 2 | 0 |
| Intrusion | 0 | 0 | 0 | 0 | 0 | 3 | 3 | 0 |

**Note.* During privacy expert evaluation (Section 5.2.3), each privacy expert was asked to select the top three risk types they saw as the most crucial to address for their assigned AI product concept.